\documentclass[11pt]{article}
\pdfoutput=1
\newcommand{\Comment}[1]{{}}
\usepackage{amsfonts,amsthm,amsmath,amssymb,slashed}
\usepackage[textwidth = 430 pt, textheight = 630 pt]{geometry}
\usepackage{color}

\Comment{\usepackage{color}
\definecolor{MyDarkBlue}{rgb}{0.15,0.15,0.45}
\usepackage[linktocpage=true]{hyperref}
\hypersetup{
colorlinks=true,
citecolor=MyDarkBlue,
linkcolor=MyDarkBlue,
urlcolor=MyDarkBlue,
pdfauthor={Horatiu Nastase and Francisco Rojas},
pdftitle={Vortices with source and nontrivial statistics in 2+1 dimensions},
pdfsubject={hep-th}
}

\usepackage[numbers,sort&compress]{natbib}
\usepackage{hypernat}}
\usepackage{graphicx}
\usepackage{cite}

\newcommand\ignore[1]{}
\def\one{{\,\hbox{1\kern-.8mm l}}}

\newcommand{\tr}{\operatorname{tr}}

\def\a{\alpha}\def\b{\beta}

\def\d{\partial}

\newcommand{\Cset}{{\,\,{{{^{_{\pmb{\mid}}}}\kern-.45em{\mathrm C}}}}}

\newcommand{\be}{\begin{equation}}
\newcommand{\bea}{\begin{eqnarray}}

\newcommand{\ee}{\end{equation}}
\newcommand{\eea}{\end{eqnarray}}

\newcommand{\eal}[1]{\be \begin{aligned} #1 \end{aligned}\ee} 
\usepackage{subfigure}

\begin{document}

\renewcommand{\thefootnote}{\fnsymbol{footnote}}

\makeatletter
\@addtoreset{equation}{section}
\makeatother
\renewcommand{\theequation}{\thesection.\arabic{equation}}

\rightline{}
\rightline{}


\vspace{10pt}


\begin{center}
{\LARGE \bf{\sc Vortices with source, FQHE and nontrivial statistics in 2+1 dimensions}}
\end{center} 
 \vspace{1truecm}
\thispagestyle{empty} \centerline{
{\large \bf {\sc Horatiu Nastase${}^a$}}\footnote{E-mail address: \Comment{\href{mailto:nastase@ift.unesp.br}}{\tt nastase@ift.unesp.br}}
{\bf{\sc and}}                                                    
{\large \bf {\sc Francisco Rojas${}^{a,b}$}}\footnote{E-mail address: \Comment{\href{mailto:frojasf@ift.unesp.br}}{\tt frojasf@ift.unesp.br}}                                    }

\vspace{.5cm}


\centerline{${}^a${\it Instituto de F\'{i}sica Te\'{o}rica, UNESP-Universidade Estadual Paulista}} 
\centerline{{\it R. Dr. Bento T. Ferraz 271, Bl. II, Sao Paulo 01140-070, SP, Brazil}}

\bigskip

\centerline{\em${}^b$Departamento de F\'isica, Facultad de Ciencias F\'isicas y Matem\'aticas, Universidad de Chile}
\centerline{\em Blanco Encalada 2008, Santiago, Chile}
\vspace{1truecm}

\thispagestyle{empty}

\centerline{\sc Abstract}

\vspace{.4truecm}

\begin{center}
\begin{minipage}[c]{380pt}
{\noindent We investigate vortex soliton solutions in 2+1 dimensional scalar gauge theories, 
in the presence of source terms in the action. Concretely, this would be applied to 
anyons, as well as the Fractional Quantum Hall Effect (FQHE). We classify solitons for renormalizable potentials, as well as some nonrenormalizable 
examples that could be relevant for the FQHE. 
The non-Abelian case, specifically for theories with global non-Abelian symmetries, is also investigated, as is the non-relativistic limit 
of the above theories, when we get a modification of the Jackiw-Pi model, with an interesting new vortex solution. 
We explore the application to the ABJM model, as well as more general SYM-CS models in 2+1 dimensions. 
}
\end{minipage}
\end{center}

\vspace{.5cm}

\setcounter{page}{0}
\setcounter{tocdepth}{2}

\newpage

\tableofcontents
\renewcommand{\thefootnote}{\arabic{footnote}}
\setcounter{footnote}{0}

\linespread{1.1}
\parskip 4pt


\section{Introduction}

Soliton solutions have played a pivotal role in the development of field theories. For instance, understanding monopoles and instantons was crucial to obtaining the first 
complete non-perturbative low energy effective action by Seiberg and Witten in the case of 4 dimensional supersymmetric gauge theories
\cite{Seiberg:1994rs,Seiberg:1994aj}. In 2+1 dimensions, vortex solutions have also played a key role in condensed matter theory, in particular 
for the theory of superconductivity, notably Abrikosov's vortex lattice. But usually, and especially in the last example (the standard 
Abrikosov-Nielsen-Olesen, or ANO, vortex), the vortices were solitonic solutions of 
the equations of motion of an abelian gauge theory involving a scalar field. 

Moreover, especially given the fact that there exists a duality between particles and vortices (in 4 dimensions there is a duality between particles and 
monopoles), which has been explained in a path integral context in \cite{Murugan:2014sfa}, one can ask whether there are relevant vortex solutions that 
need a source term, just like the electron is a source for electromagnetism. In the case of string solitons of $p$-brane type \cite{Duff:1994an},
it is well known that a source term is needed for both "electric" and "magnetic" solutions, albeit for magnetic solutions it drops out of the calculations. 
More relevantly, one way to describe an anyonic particle (with fractional statistics) is to add a delta-function magnetic field $B=F_{12}$ associated with 
the particle position, and such a magnetic field source  can be obtained from adding a source term and a Chern-Simons term for the gauge field. 
In such a way anyons can be obtained in the Fractional Quantum Hall Effect, via an effective action of the Chern-Simons type, and also 
by an analysis of the possible wavefunctions. 

It is then a relevant question to ask whether we can find more general vortex solutions in 3 dimensional field theories that have Chern-Simons
terms, by including source terms for the vortices.\footnote{Note that vortex solitons in Chern-Simons theories, relevant to condensed matter
theories and the FQHE, have been considered before, for instance in \cite{Tafelmayer:1993ag}, see \cite{Horvathy:2008hd} for a review, 
but here we consider them in the presence of a {\em source}, with a different ansatz. We believe this is a novel approach.} 
This has been considered briefly in \cite{Murugan:2014sfa}, but here we will systematically 
explore the issue.  We will carefully study the case of a general renormalizable potential, and find under what conditions do we have a solution. 
We will then consider some examples of nonrenormalizable potentials that have relevance to the physics of the FQHE. 
We will then consider the case of a potential for a scalar field with a non-Abelian global symmetry. Moreover, since the nonrelativistic limit of Chern-Simons-scalar theories is known to be relevant in condensed matter systems, we consider the non-relativistic limit of the models we have found, and find a very 
interesting new type of nonrelativistic vortex solution. We will finally see if we can embed the soliton solutions that we have found in interesting CS+scalar models in 2+1 dimensions, specifically the 
ABJM model and SYM-CS systems. 

The paper is organized as follows. In section 2 we consider the motivation coming from condensed matter, namely anyonic physics and the FQHE. 
In section 3 we examine the generality of solutions of abelian models. In section 4 we generalize to non-Abelian models, and in section 5 we 
take the non-relativistic limits. In section 6 we embed in supersymmetric CS theories, and in section 7 we conclude.

\section{Condensed matter motivation}

The motivation for the present work comes from condensed matter physics, thus in this section we present the implications of 
having delta function sources together with Chern-Simons terms.

\subsection{Anyons}

The explanation of why adding a delta function magnetic field situated at the position of a particle makes that particle anyonic is a standard one
(see for instance section 2.1 in \cite{Dunne:1998qy}, as well as \cite{Rao:1992aj,stern}).

Consider an action for a Chern-Simons field coupled to a (vortex) current,
\be
S=\int d^3x \left[-\frac{k}{4\pi}\epsilon^{\mu\nu\rho}A_\mu\d_\nu A_\rho+\frac{1}{e}A_\mu j^\mu_{\rm vortex}\right]\;,
\ee
where the (vortex) current is, in the static case, simply a delta function,
\be
j^0_{\rm vortex}=\delta(z-z_0(t)).
\ee
Then the equations of motion are
\be
\epsilon^{\mu\nu\rho}F_{\nu\rho}=\frac{4\pi}{ke}j^\mu_{\rm vortex}(t)\Rightarrow B\equiv F_{12}=\frac{2\pi}{k e}\delta(z-z_0(t))\;,\label{anyon}
\ee
and we see that we have a magnetic field localized only at the position of the delta function source.  
In the gauge $A_0=0$ the equations of motion have the solution
\be
A_i=\frac{1}{ek}\d_i arg(\vec{x}-\vec{x}_0)\;,
\ee
which generalizes to the case of $N$ separated sources $\sum_i\delta(z-z_i(t))$ as 
\be
A_i=\frac{1}{ek}\sum_{a=1}^N \d_i arg(\vec{x}-\vec{x}_a)=\frac{1}{ek}\sum_{a=1}^N\epsilon^{ij}\frac{(x^j-x^j_a(t))}{|\vec{x}-\vec{x}_a(t)|^2}.
\ee
Except at the positions of the sources (vortices), this is a pure gauge, but removing it by a gauge transformation would imply that a scalar field or 
wavefunction $\psi(\vec{x})$  would acquire a phase, 
\be
\psi(\vec{x})\rightarrow \exp\left[-i\frac{1}{k}\sum_{a=1}^N arg(\vec{x}-\vec{x}_a)\right]\psi(\vec{x}).
\ee
The Hamiltonian of $N$ such particles (or rather, vortices, see later) in the presence of the singular gauge field $A_i$ would be
\be
H=\frac{1}{2m}\sum_{a=1}^N|\vec{p}_a-e\vec{A}(\vec{x}_a)|^2\;,
\ee
where
\be
A_i(\vec{x}_a)=\frac{1}{ek}\sum_{b\neq a}^N\epsilon^{ij}\frac{(x_a^j-x_b^j)}{|\vec{x}_a-\vec{x}_b|^2}.
\ee

Moving one particle around another results then in an Aharonov-Bohm phase
\be
\exp \left[ie\oint_{C=\d S} \vec{A}\cdot d\vec{x}\right]=\exp\left[ie\int_S F_{12}dx^1\wedge dx^2\right]=\exp\left[\frac{2\pi i}{k}\right]\;,
\ee
which can be interpreted as a double exchange of two identical particles, thus having an anyonic exchange phase of 
\be
\theta=\frac{\pi}{k}.
\ee
In conclusion, coupling the Chern-Simons term with a  source results in anyonic particles at the position of the source. 
We want these particles to actually be vortices in a more general theory, in which we embed the Chern-Simons plus source term.

The gauge field above, obtained from a gauge transformation, is called a {\em statistical gauge field}, and is an emergent one. 
Indeed, the Chern-Simons action that describes it has no degrees of freedom. Note however that for the anyon argument it does not 
matter whether $A_\mu$ is the emergent or the electromagnetic gauge field, all we need is the presence of the gauge field flux, which will 
lead to the Aharonov-Bohm phase giving fractional statistics. 

\subsection{The Fractional Quantum Hall Effect}

We can in fact have anyons in the Fractional Quantum Hall Effect. An effective action that describes the FQHE is 
written in terms of the electromagnetic field $A_\mu$ and the statistical (emergent) gauge field $a_\mu$ as (see for instance 
\cite{Witten:2015aoa}, and the original reference \cite{Zhang:1988wy})
\be
S_{\rm eff}=\int d^3x \left[-\frac{r}{4\pi}\epsilon^{\mu\nu\rho}a_\mu \d_\nu a_\rho+\frac{1}{2\pi} \epsilon^{\mu\nu\rho}A_\mu\d_\nu a_\rho\right]
+q\int dt a_0(x_0,t).
\ee
In the absence of the source term, we could integrate out $a_\mu$ via its equation of motion,
\be
f_{\mu\nu}=\frac{1}{r} F_{\mu\nu}\;,
\ee
which means that, up to a gauge transformation, we have $a_\mu=A_\mu/r$, and by replacing back in the action we get
\be
S'_{\rm eff}=\frac{1}{r}\frac{1}{4\pi}\int d^3x \epsilon^{\mu\nu\rho}A_\mu \d_\nu A_\rho.
\ee
The appearence of the fractional coefficient $1/r$ instead of the integer one $k$ suggests a fractional quantum Hall conductivity, which is indeed the case.

The charge carriers added to the effective action
are anyonic quasi-particles, with integer charge $q$ under the statistical gauge field $a_\mu$. We can also add 
an explicit electric charge coupling to the effective action,
\be
e\int dt A_0.
\ee
The equation of motion for $a_0$ is 
\be
\frac{F_{12}}{2\pi}-\frac{r f_{12}}{2\pi}+q\delta(x-x_0)=0\;,
\ee
which can be solved by either $F_{12}$ or $f_{12}$ being a delta function.

If one chooses $f_{12}$, which is a truly 2+1 dimensional field (statistical gauge field), to be a delta function,
\be
\frac{f_{12}}{2\pi}=\frac{q}{r}\delta(x-x_0)\;,
\ee
then by substituting it back in the effective action we get an effective electric charge coupling (from the coupling to $A_0$ of $f_{12}$) of 
\be
J_0=\frac{q}{r}\delta(x-x_0)\;,
\ee
i.e., these quasi-particles have fractional electric charge. 

But we have also another possibility.
Since $F_{\mu\nu}$ is a 3+1 dimensional field restricted to 2+1 dimensions, we can solve by $F_{12}$ being the delta function, 
by considering the extension in the third spatial dimension, specifically a flux tube in it. In the absence of vortices, this 
interpretation is shaky, since we could only consider small current loops, that create the above flux tube in the center, but that 
also needs corresponding anti-fluxes nearby, generated by opposite loops. However, a vortex obtained {\em when adding scalar fields} 
to the system makes more sense: it will have the flux at the core, and the only constraint is to have an equal number of vortices and anti-vortices 
on the plane, so that there is no total flux escaping in the third spatial dimension. 

Vortices also are necessary for the standard explanation of the FQHE by Robert Laughlin, using Laughlin's wavefunction
(postulated as an approximation to the true ground state, but verified experimentally to be approximately correct 
to a very high degree of accuracy) for $N$ electrons at $\vec{r}_i$, $i=1,...,N$, of 
\be
\Psi_m(\vec{r}_1,...,\vec{r}_N)=C_{N,l,m}\prod_{i<j=1}^N(z_i-z_j)^m e^{-\sum_{i=1}^N |z_i|^2/4l^2}.
\ee
Here $m$ is odd, $m=2p+1$, and this wavefunction corresponds to the FQHE at filling fraction (ratio of filled Landau levels) of 
$\nu=1/m$.  For $N=1$ (a single electron), we have
\be
\Psi_m(z)=z^m e^{-|z|^2/4l^2}\label{laughlin}
\ee
which corresponds to $m$ vortices situated at $z=0$ ($\psi(z)=z^m$ in polar coordinates becomes $\psi(r,\theta)=r^m e^{im\theta}$). 

On top of this ground state, the state with a "quasi-hole" at the origin is 
\be
\Psi=\left(\prod_i z_i\right)\Psi_m=\left(\prod_i |z_i|\right)e^{-\sum_i \phi_i}\Psi_m.
\ee
Thus the quasi-hole has a quantum of vorticity (it is a vortex), has an electric charge $\nu=1/m$, and putting $m$ of them together makes up 
an electron, in the Laughlin ground state.

Thus a vortex with a scalar field profile, with a source term, could generate the real magnetic field $F_{12}$ at the origin, or the $f_{12}$ magnetic 
field, and could be identified with the quasi-hole needed by Laughlin. The scalar field can be thought of as a composite field in terms of the 
fundamental degrees of freedom.

\section{Abelian vortex solutions}

We have now defined the starting point of our analysis: we want a scalar field coupled to a gauge field, including a Chern-Simons term, and 
with a source term in the action. We will analyze the possible vortex solutions of such an action.
In fact, the analysis of abelian vortex solutions with source was begun in \cite{Murugan:2014sfa}, but most of the analysis was missed there, so 
here we will do the full analysis.

\subsection{Set-up and general analysis}

With the idea of being able to apply to the FQHE, we consider an effective action for the statistical field $a_\mu$, electromagnetic 
field $A_\mu$ and a complex scalar field $\Phi$,
\be
S=\int d^3x \left[-\frac{1}{2}|(\d_\mu-iea_\mu)\Phi|^2 -\frac{k}{2\pi}\epsilon^{\mu\nu\rho}a_\mu\d_\nu A_\rho +\frac{1}{e}A_\mu j^\mu_{\rm vortex}(t)
-V(|\Phi|)\right]\;,\label{action}
\ee
where for the FQHE we would put $k=1$, and possibly with a Chern-Simons term for the statistical gauge field and a source for it, making it 
an anyon,
\be
S'=\int d^3x \left[-\frac{r}{4\pi}\epsilon^{\mu\nu\rho}a_\mu \d_\nu a_\rho+\frac{q}{e}a_\mu j^\mu_{\rm vortex}\right].
\ee
The equations of motion of $S$, after imposing $A_\mu=0$ (zero electromagnetic field), are
\bea
\epsilon^{\mu\nu\rho}\d_\nu a_\rho&=&\frac{2\pi}{ke} j_{\rm vortex}^\mu\cr
\Phi(D_\mu\Phi)^*-\Phi^*(D_\mu\Phi)&=&0\cr
(D_\mu)^* D^\mu\Phi&=& \frac{dV}{d\Phi}\;,
\eea
where $D_\mu=\d_\mu-iea_\mu$. Adding $S'$ to $S$ would only change the equation for $a_\mu$ (the second one above), by 
adding a term $-\frac{r}{2\pi}\epsilon^{\mu\nu\rho}\d_\nu a_\rho+\frac{q}{e}j^\mu_{\rm vortex}$.

Note that we consider $A_\mu=0$ (zero electromagnetic field) since we are now interested in a solution that doesn't have its topology associated 
with a nontrivial magnetic field $\vec{B}$, like for the usual ANO vortex, but rather it's associated to a nontrivial statistical gauge field. 
Indeed, there is no evidence for a nontrivial ANO-type vortex structure in the FQHE, like in the case of the Abrikosov vortex lattice for superconductors. 

We consider the usual one-vortex ansatz,
\be
\Phi(r,\theta)=|\Phi(r)| e^{i\theta}\;,
\ee
where $\theta$ is the polar angle, satisfying 
\be
\epsilon^{0\mu\nu}\d_\mu\d_\nu\theta=2\pi \delta(z).
\ee
The equations of motion for $S$ then become
\bea
\epsilon^{\mu\nu\rho}\d_\nu a_\rho&=&\frac{2\pi}{ke} j_{\rm vortex}^\mu\cr
\Phi(D_\mu\Phi)^*-\Phi^*(D_\mu\Phi)&=&0\cr
\frac{|\Phi(r)|''}{|\Phi(r)|}&=&2\frac{d V(|\Phi|)}{d|\Phi|^2}\;,\label{eom}
\eea
Note that with respect to  \cite{Murugan:2014sfa}, now the first equation in (\ref{eom}) has an explicit source on the right hand side. 

We consider also an ansatz for the gauge field that solves the 
second equation of motion explicitly. We see that $D_\mu \alpha=0$, where $\alpha $ is the argument of $\Phi$ ($\Phi=|\Phi| e^{i\a}$),  i.e.
\be
e a_\mu=\d_\mu\theta\;,
\ee
solves it, and it also solves the first equation of motion everywhere, including at $r=0$, if (and only if) $k=1$. Therefore if $k=1$ we 
can consider all the equations of motion to be satisfied at $r=0$, whereas if $k\neq 1$, we consider that we have an idealization of some 
real situation, and the equation of motion at $r=0$ need not be satisfied. 

Adding $S'$ to $S$, the extra term in the $a_\mu$ equation of motion vanishes on the ansatz $ea_\mu=\d_\mu \theta$, 
$\epsilon^{\mu\nu\rho}\d_\nu \d_\rho \theta=\frac{2\pi}{ke} j^\mu_{\rm vortex}$, if $r/q=k$ (so, if $k=1$, for $r=q$). 

Since the equation for the gauge field ansatz can be written as $\d_\theta \a=a_\theta$, 
and is also valid at infinity, this gives the usual charge quantization condition $\oint d\theta a_\theta =\oint d\a$, so $\a=\theta$. 
Note however that the charge that is quantized is the topological {\em statistical charge} associated with $a_\mu$, and not the magnetic
charge associated with $A_\mu$. The solution has statistical magnetic flux $f_{ij}$, not usual magnetic flux $F_{ij}$.

Finally then, one is left with an equation of motion for $|\Phi(r)|$ to solve in order to find the vortex solution. The equation,
\be
|\Phi(r)|''=\frac{dV(|\phi|)}{d|\Phi|}\;,
\ee
takes the form of a classical mechanics problem for motion in an inverted potential $V_m(x)=-V(x)$,
\be
\ddot X(t)=-\frac{dV_m(X)}{dX}\;,
\ee
where time $t$ is replaced by radius $r$ and position $X$ is replaced by field $|\Phi|$. 

The usual argument for vortices, that we need $|\Phi(r=0)|=0$ in order for the solution to make sense (since it gets multiplied by the 
phase $e^{i\theta}$ that is ill defined at $r=0$), still holds. 

Less clear is the case of the usual condition for the one-vortex solution, $|\Phi(r)|\sim A r$ as 
$r\rightarrow 0$ (imposed such that $|\Phi(z)|\sim z$, where $z=r e^{i\theta}$), 
and for an $N$-vortex solution $|\Phi(r)|\sim A_N r^N$ (so that $\Phi(z)\sim z^N$). We will see that now it makes sense to consider 
more general possibilities for the behaviour near $r\sim 0$. 
This analysis however would all be valid only for the $k=1$ case; 
otherwise the equations of motion at $r=0$ are not satisfied anyway, so we need to think of the system as an idealization of some 
real, but smooth, system, that will not need to satisfy the equation at $r=0$. 

\begin{figure}[ht!]
\centering{
\subfigure[]{
\includegraphics[width=.42\textwidth]{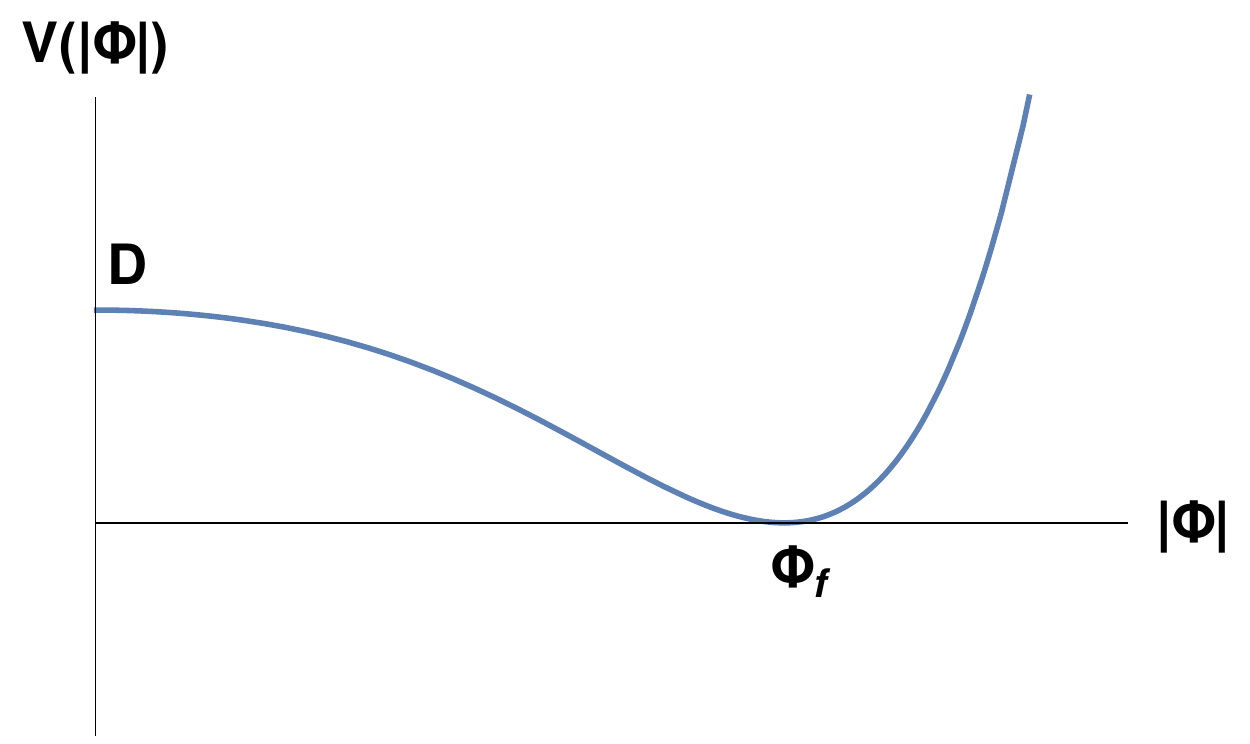}
\hspace{1cm}}       
\subfigure[]{
\includegraphics[width=.42\textwidth]{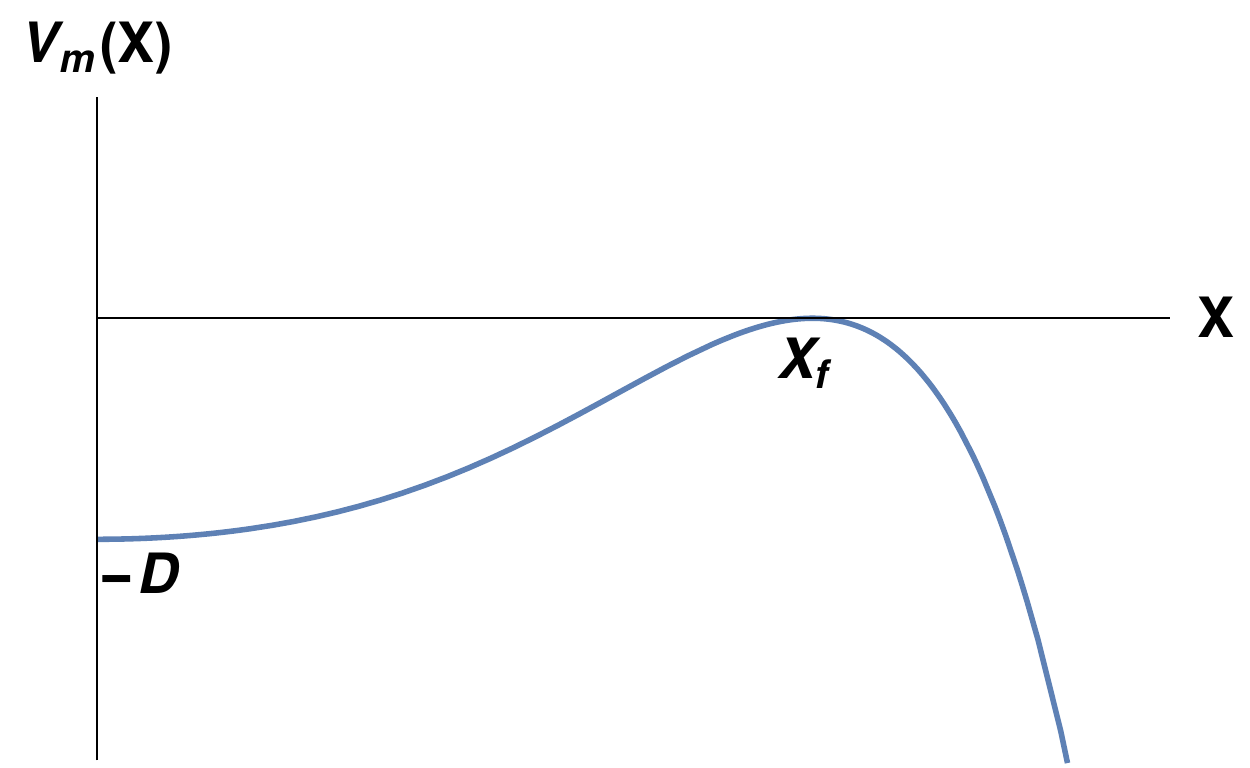}}
}
\caption{\small Depicted in (a) the potential admitting a vortex solution of the form $|\Phi| \sim A r$ for $r \sim 0$. This needs $m^2<0$ with $C_1>0$, 
$\forall \lambda$ or $\lambda >0$ and $C_1=0$. In (b) the mechanical analogy is shown: A particle is sent up a frictionless hill with some initial 
velocity, just large enough to approach the location $X_f$ asymptotically as $t \to \infty$. 
\label{fig:nonzerovim2neg}}
\end{figure}

\begin{figure}[ht!]
\centering{
\subfigure[]{
\includegraphics[width=.42\textwidth]{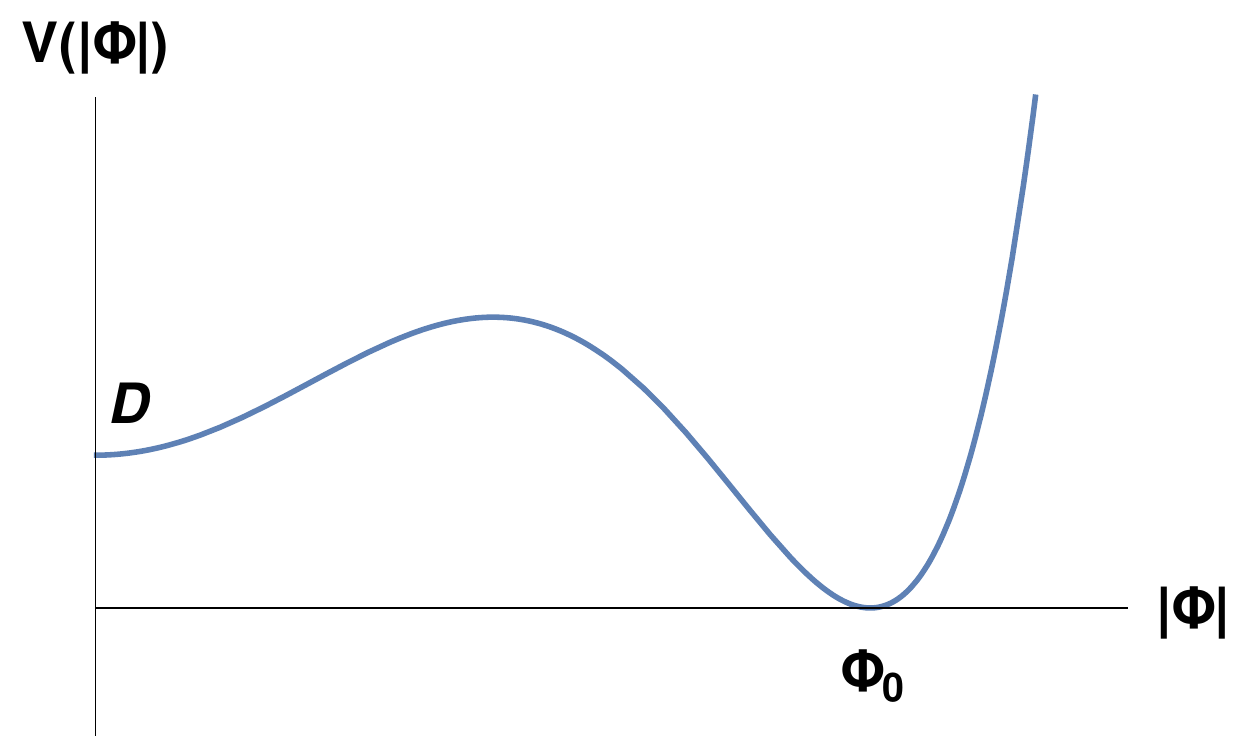}
\hspace{1cm}}       
\subfigure[]{
\includegraphics[width=.42\textwidth]{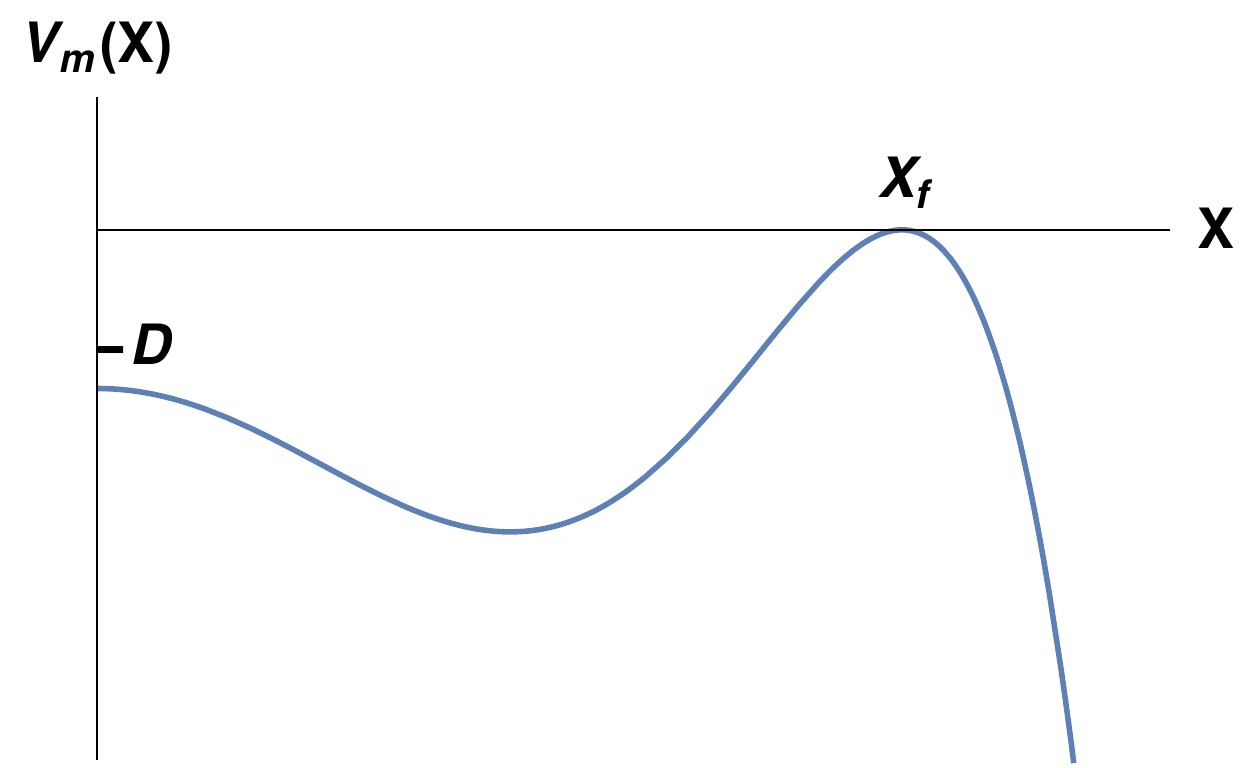}}
}
\caption{\small Depicted in (a) the \emph{sextic} potential admitting a vortex solution of the form $|\Phi| \sim A r$ for $r \sim 0$. Here we need 
$m^2>0$, $\lambda<0$, and $C_1>0$. In (b) the mechanical analogy corresponds to a particle shot down a hill with some initial velocity which 
is just right to climb up the hill and asymptotically approach $X_f$ as $t \to \infty$. 
\label{fig:nonzerovim2pos}}
\end{figure}

The picture of a classical (point mass) motion in an upside down potential allows us to get a simple intuition for whether there are solutions 
or not. The motion in $V_m(X)$ is frictionless, so energy (kinetic plus potential) is conserved. We start off at $t=0\rightarrow r=0$ with 
a position $X(t=0)\rightarrow |\Phi(r=0)|$ and a velocity $dX/dt(t=0)\rightarrow d|\Phi|/dr(r=0)$, and the question is whether the point mass 
can come to rest at $t=\infty \rightarrow r=\infty$ with zero velocity $dX/dt(t=\infty)=0\rightarrow d|\Phi|/dr(r=\infty)=0$ (only potential energy, 
no kinetic). That is only possible if the potential $V_m(x_f)\rightarrow -V(|\Phi_f|)$ has an extremum ($dV/d|\Phi|=0$) at the position $|\Phi_f|$ 
where we end up at infinity. See Figures~\ref{fig:nonzerovim2neg} and~\ref{fig:nonzerovim2pos} showing the relevant cases for our analysis.

For the condition at $r=0$, we see that $|\Phi|\sim r^a$, with $a<1$ is excluded, since then the velocity at zero, $d|\Phi|/dr(r=0)\rightarrow \infty$, so 
motion with infinite initial kinetic energy can only lead to infinite final potential energy, which is impossible (and contradicts the finite energy 
condition for the soliton). 

If $|\Phi(r)|\sim r$, then we have a finite velocity at zero, thus a finite initial kinetic energy for motion in $V_m(X)$, 
which means that the final point $X_f\rightarrow |\Phi_f|$ can only be at a higher value of $V_m(X)$, which means a lower value 
of $V(|\Phi|)$. There are basically two classes of potentials admitting this type of vortex solutions characterized by the sign of $m^2$. 
Figures~\ref{fig:nonzerovim2neg} and Figure~\ref{fig:nonzerovim2pos} show the potentials for $m^2<0$ and $m^2>0$ respectively, denoted
{\bf type 1} and {\bf type 2} in the following. Note that for $m^2>0$ (depicted in Figure~\ref{fig:nonzerovim2pos}) the potential needs to be sextic.

Finally, if $|\Phi(r)|\sim r^p$, with $p>1$, the velocity at zero vanishes, $d|\Phi|/dr(r=0)=0$ (though there can be acceleration, $d^2|\Phi|/dr^2\neq 0$, 
if $p=2$), 
which means we move in $V_m$
with zero initial kinetic energy, thus the initial and final points must have the same value of $X\rightarrow |\Phi|$, i.e. 
$|\Phi_0|=|\Phi(r=0)|=|\Phi(r=\infty)|=|\Phi_f|$. We will however see (below eq. (\ref{phirzero})) that this possibility doesn't satisfy the equations of motion.  
In all cases $|\Phi_f|$ must be an extremum of $V(|\Phi|)$. 

We must also consider that we want finite energy solitons. Because of this, the extremum where $\Phi$ ends up at $r\rightarrow \infty$ 
must have $V(|\Phi_f|)=0$, otherwise the constant energy density would integrate to infinity. 

We can then easily summarize the condition for a potential to admit this kind of solitons. $V$ must have an extremum ($dV/d|\Phi|=0$) at
a nonzero $|\Phi_f|\neq 0$, and we need $V(|\Phi|<|\Phi_f|)\geq 0$ (so $V_m(X<X_f)\leq 0$, since the point mass must have smaller
potential energy than at infinity throughout its trajectory).

Note that in all these solutions, we have $|\Phi(r=0)|=0$, which then implies that we need $|\Phi(r=\infty)|\neq 0$. Indeed, if the point 
mass moving in $V_m$ leaves $X=0$ at $t=0$ and reaches some $X_m$ at finite time, after which it comes back towards $X=0$, it 
will necessarily reach $X=0$ again in finite time (corresponding to finite $r$ for $|\Phi(r)|$), so this possibility is excluded.

{\bf Compactons}

\begin{figure}[ht!]
\centering{ 
\subfigure[]{
\includegraphics[width=.4\textwidth]{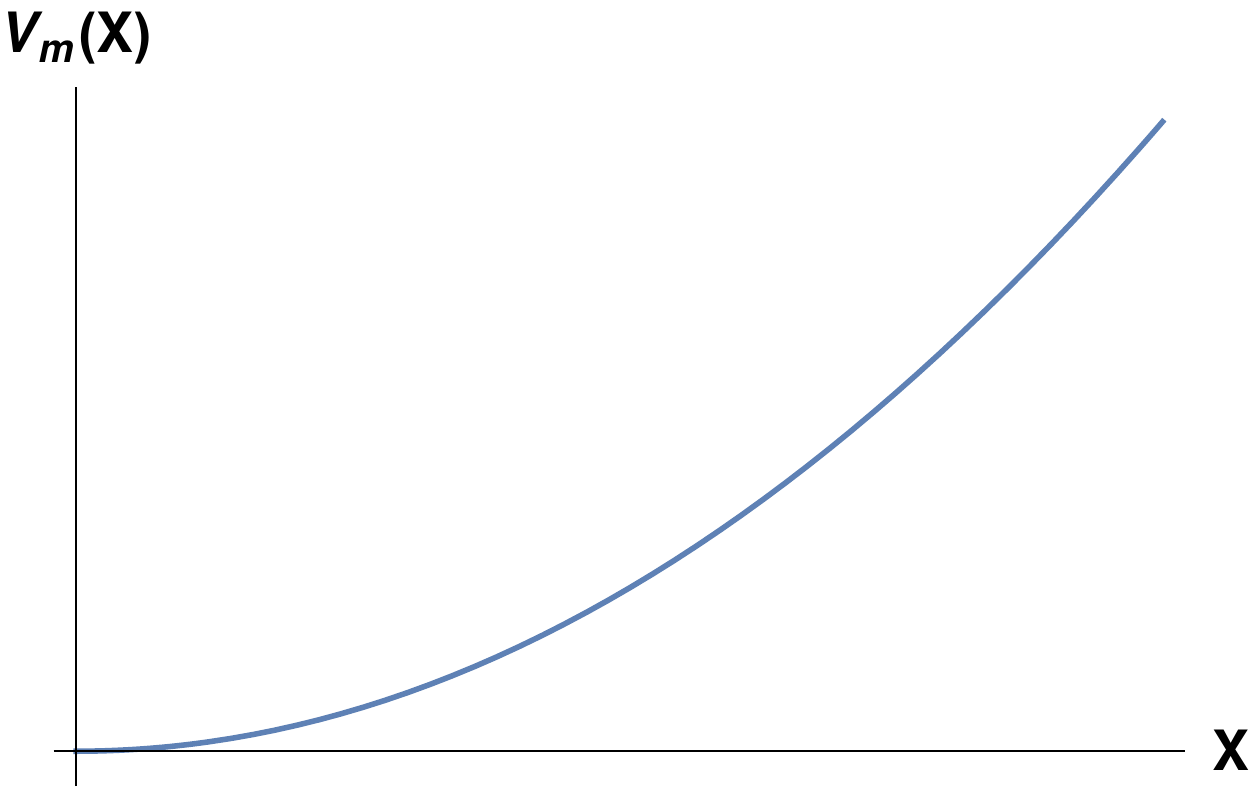}
\hspace{1cm}}
\subfigure[]{
\includegraphics[width=.4\textwidth]{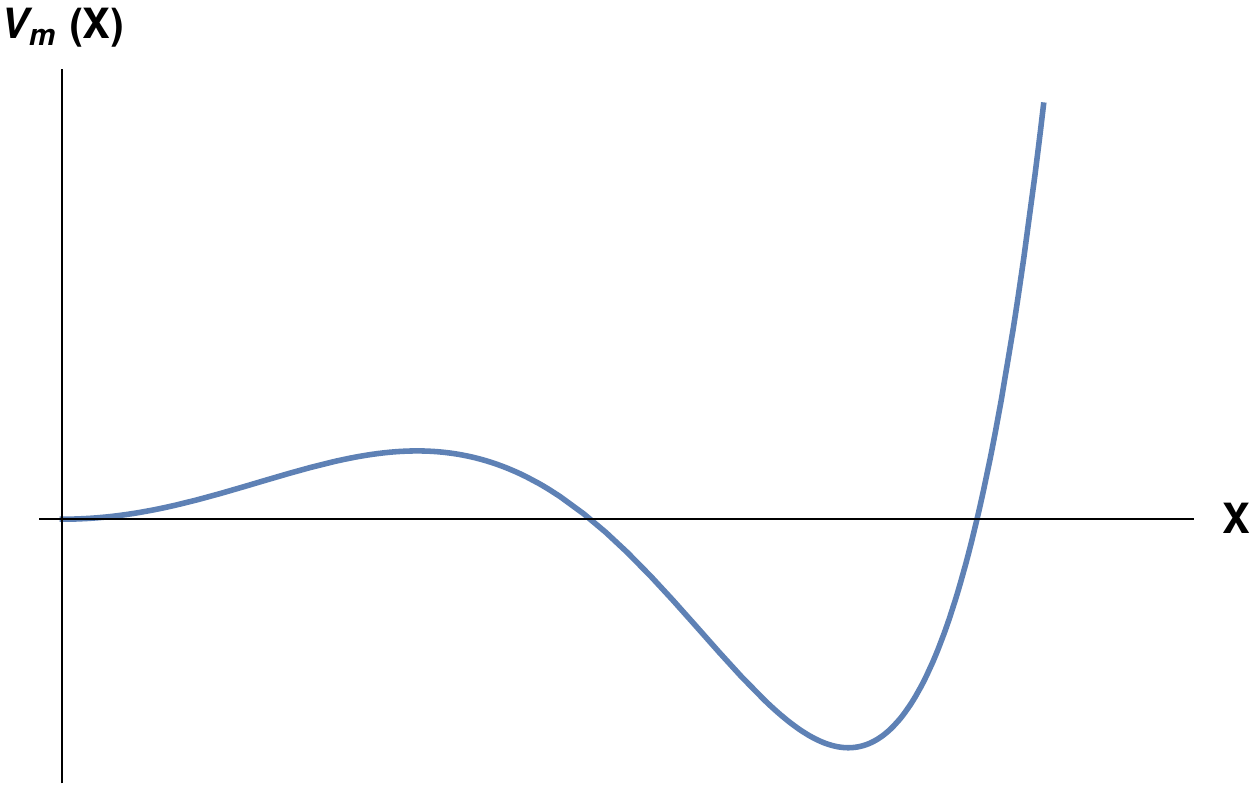}}
}\\
\centering{ 
\subfigure[]{
\includegraphics[width=.4\textwidth]{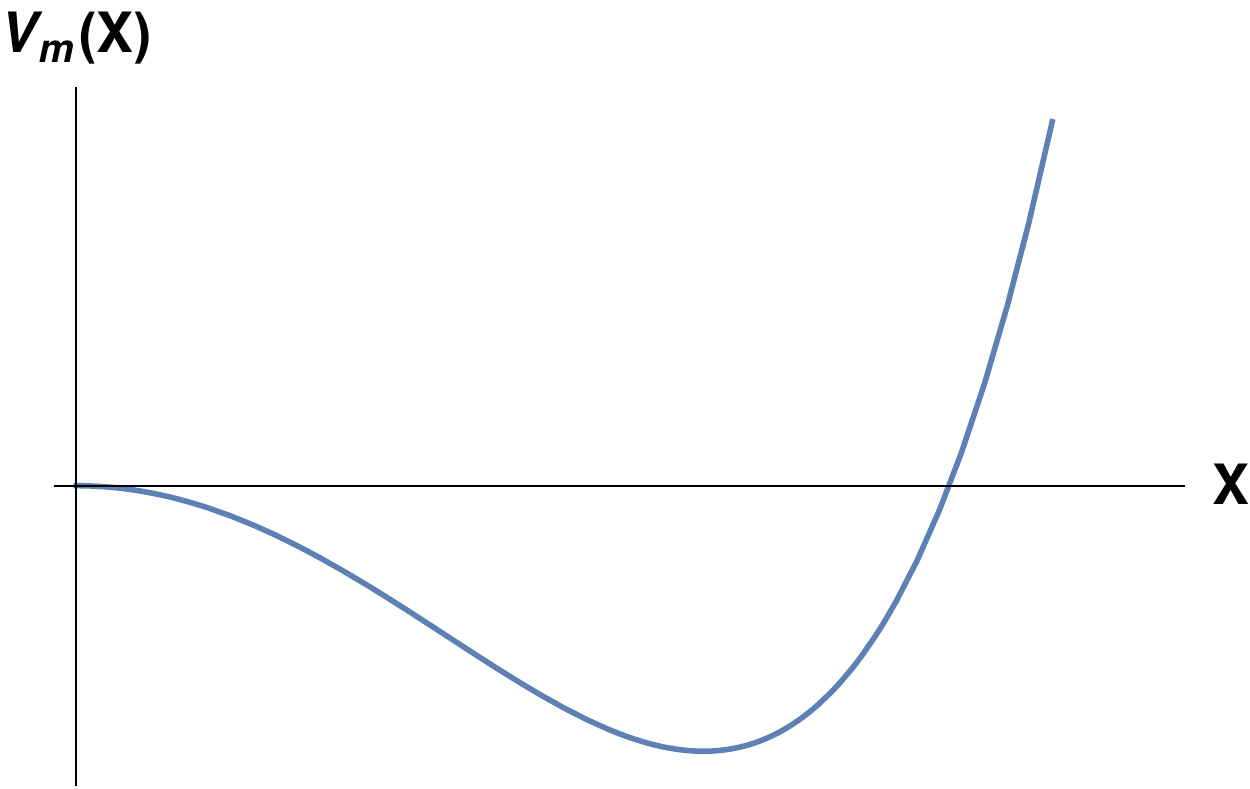}
\hspace{1cm}}
\subfigure[]{
\includegraphics[width=.4\textwidth]{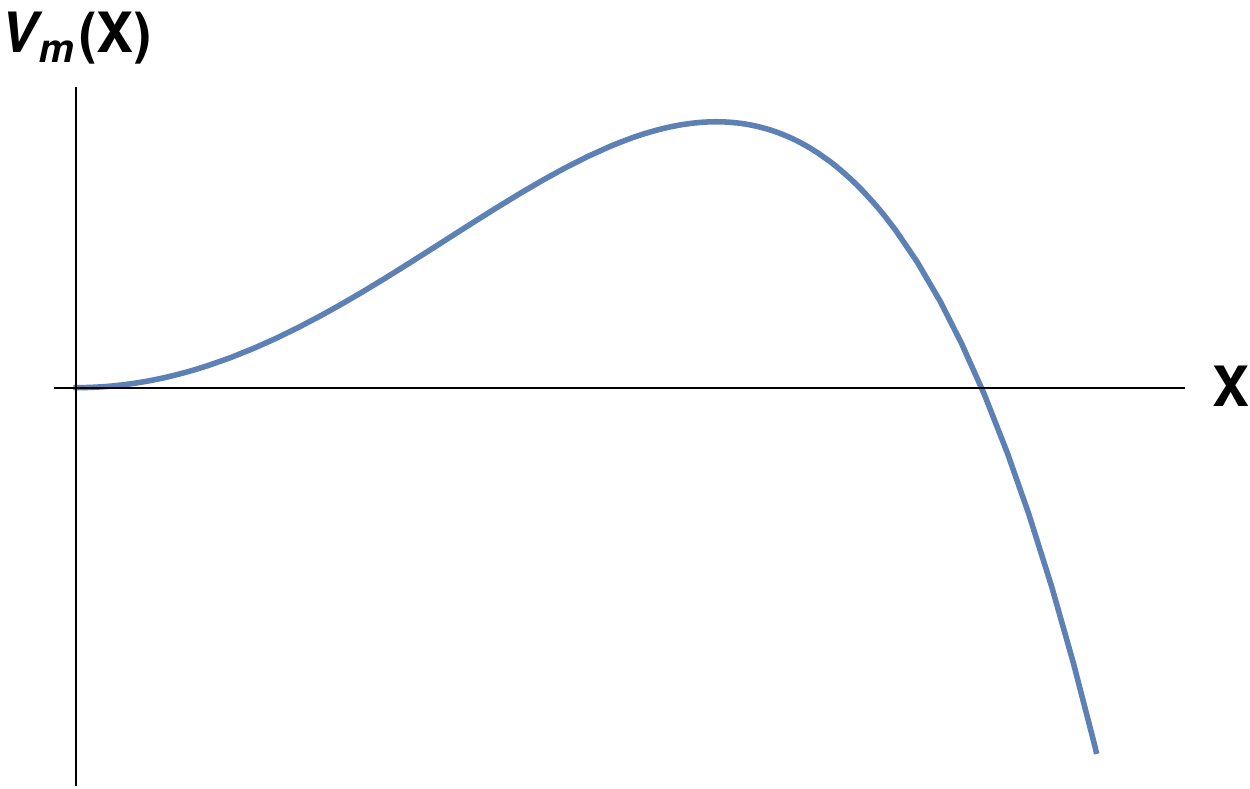}}
}
\centering{ 
\subfigure[]{
\includegraphics[width=.4\textwidth]{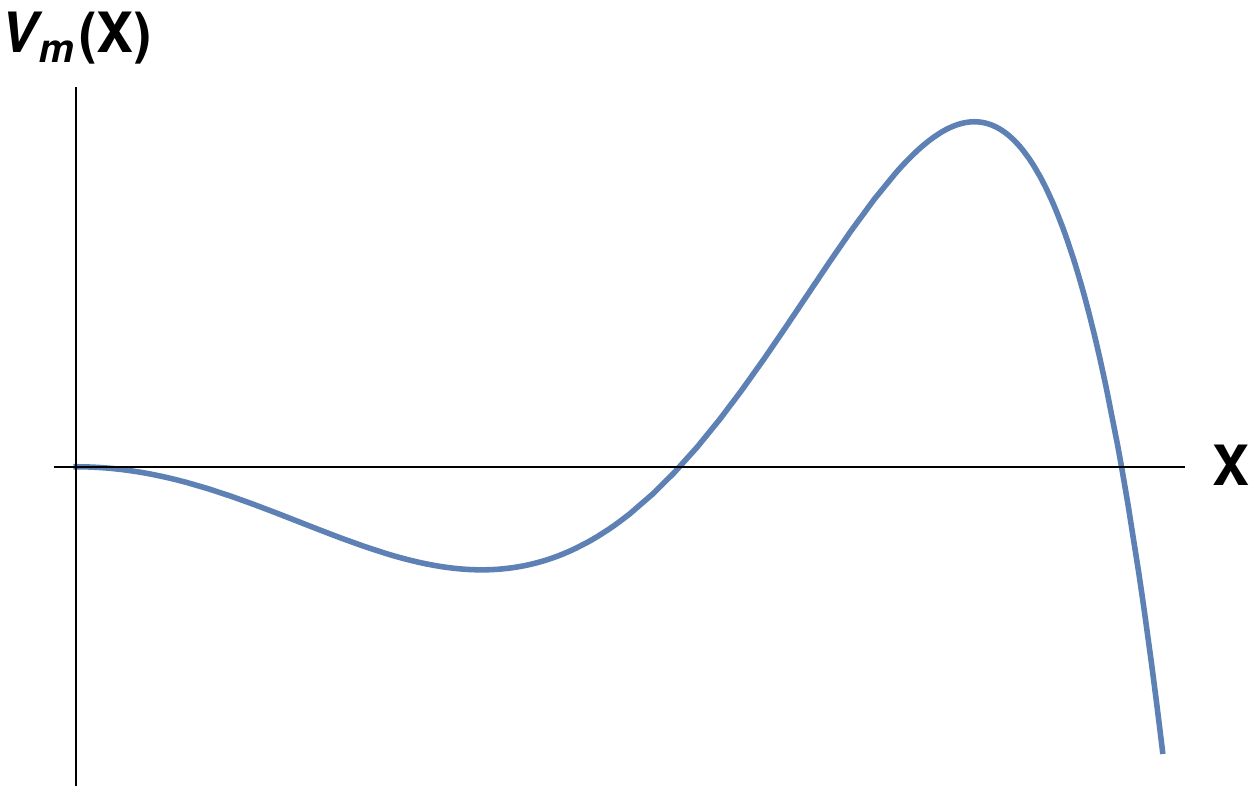}
\hspace{1cm}}
}
\caption{\small Shapes of the mechanical analog for all possible potentials admitting compacton vortex solutions. In all of these cases the 
particle starts off at $X=0$ with some initial velocity. Note that for the potentials in (a), (b) and (c) there is no limit for how large the initial 
velocity can be, since the particle will always come back to $X=0$ after some finite time. However, for the potentials shown in (d) and (e), 
the initial velocity must not be too large in order not to ``fall'' to the other side of the ``hill'', in which case the motion continues eternally 
for $t \to \infty$ (and thus the vortex does not have compact support). Note also that while the potentials shown in (a), (c) and (d) are 
not necessarily sextic, the ones in (b) and (e) are.
\label{fig:compactons}}
\end{figure}

But in the above analysis we have assumed that we reach $r\rightarrow \infty$. However, that is actually not necessary, we can 
have solutions with finite support, so called ``compactons'' \cite{Rosenau:1993zz}.\footnote{These solitons are solutions of a 
generalization of the KdV equation (see \cite{Rosenau:1993zz} for details).} These would be solutions for which $\Phi=0$ for $r\geq r_1$. 
Since $|\Phi|$ must be continuous, we need that $|\Phi(r=r_1-\epsilon)|=0$ as well, and since as we saw  we wanted $|\Phi(0)|=0$, 
the classical motion in $V_m=-V$ must be symmetrical: we start off at a $X=0$, perhaps with some velocity, then move up to some 
$X_m\rightarrow |\Phi_m|$, and then come back to $X=0$ at $t=t_1\rightarrow r=r_1$. The possibilities are plotted in Figure~\ref{fig:compactons}.
The cases in figures (a), (b) and (d), with negative leading term, will be called {\bf type 3} in the following, and cases in figures (c) and (e), with 
positive leading term in the potential $V(|\Phi|)$, will be called {\bf type 4}. Note that for all these cases also $|\Phi(r)|\sim A \; r$ at $r\rightarrow 0$, 
corresponding to motion with an initial velocity. The case of $|\Phi(r)|\sim A r^p$, with $p>1$, will again be excluded by the equations of motion.

Note that the fact that $d|\Phi|/dr$ (or a higher order derivative) is discontinuous at $r=r_1$ is not a problem. The only thing we need is that 
the equations of motion at $r=r_1$ are satisfied, which means that we need that the second derivative is continuous,
\be
0=\frac{d^2|\Phi|}{dr^2}(r_1)=\frac{dV}{d|\Phi|}(|\Phi|=0).
\ee
This condition means that we need that $|\Phi|=0$ is an extremum of $V$. 

Note also that, since the field has compact support, any (non-divergent) solution will have finite energy upon integrating over $r$. Thus, as opposed to the previous cases, there is no need to impose extra conditions on $V$.

\subsection{Renormalizable potentials}

We now consider various possible potentials and analyze the solutions in more detail. 
We start the analysis with renormalizable scalar potentials, having in mind that the composite scalar is nevertheless a quantum field. 
We will treat the case that $V$ comes from a full quantum effective action, so can be nonrenormalizable, in the next subsection.

In 3 dimensions, the most general renormalizable potential is sextic. Since moreover, we want the potential to only depend on $|\Phi|^2=\Phi^\dagger
\Phi$, we consider the general potential
\be
V(|\Phi|)=C_1|\Phi|^6+\lambda |\Phi|^4+m^2|\Phi|^2+D\;,
\ee
giving the equation of motion 
\be
\frac{|\Phi|''}{|\Phi|}=\frac{dV}{d|\Phi|^2}=m^2+2\lambda|\Phi|^2+3C_1|\Phi|^4.
\ee

In the following we will assume that $D$ is such that the solution ends up in an extremum point ($dV/d|\Phi|=0$) with $V=0$, and we will 
not write it explicitly. This is needed, since we want to have a solution with finite energy, otherwise, if $V\neq 0$ at $r\rightarrow \infty$, we would get 
an infinite contribution to the energy.

We will treat the potential case by case. An important condition is whether $m^2$ is positive (a mass term) or negative (an instability), 
so we will split the analysis according to this condition.

{\bf I \underline{$m^2\geq 0$.}}

If $m^2>0$, $V_m=-V$ starts going down at $|\Phi|=0$, and is a local maximum for $V_m$. Therefore we need to be in type 2 for 
the usual solitons, or type 4 for the compactons. Note that all of them require that the potential $V$ starts to go down after the initial upward 
trend, so we need either $\lambda<0$ or $C_1<0$, but both cannot be positive (nor zero). 

{\bf A. \underline{$C_1\neq 0, \lambda\neq 0, m^2\neq 0$}. }

We start off with the solution of the type 1, that has a nonzero velocity at $r=0$, i.e. $|\Phi(r)|\sim A r$ as $r\rightarrow 0$. 

Near the origin, $r\rightarrow 0$, considering the first power law correction to the vortex behaviour $\sim A r$, i.e. 
\be
|\Phi|\sim Ar+Cr^p+...\;,
\ee
the equation of motion fixes $p=3$ and $C=Am^2/6$, so we have
\be
|\Phi|\sim Ar\left(1+\frac{m^2r^2}{6}\right).\label{phirzero}
\ee

We could imagine a solution with $|\Phi(r)|\sim A r^N$, with $N>1$, as $r\rightarrow 0$. But that doesn't satisfy the equation of motion, since 
then $|\Phi|''\sim N(N-1)r^{N-2}$, whereas $dV/d|\Phi|\sim r^N$ or higher, independent of whether $C_1,\lambda $ or $m^2$ are zero. So a solution of 
this type  is actually never possible, because of the equations of motion. In fact, the same argument shows that a compacton solution of this type is also 
always excluded.

We will only consider the compacton solutions at the end of this subsection, so we will ignore them for now. 

{\em Solution with massless "Higgs"?}

Near $r\rightarrow \infty$, in \cite{Murugan:2014sfa} it was considered a power law behaviour, 
\be
|\Phi|\sim \tilde A+\frac{\tilde B}{r^n}+...
\ee
Then the equations of motion lead at first order to 
\be
|\Phi_0|^2\equiv \tilde A^2=-\frac{m^2}{\lambda}\,,\,\,\,\,\,\, C_1=\frac{\lambda^2}{3m^2}\,.
\ee
Note that $\tilde A^2=-m^2/\lambda$, with $m^2>0$ implies $\lambda<0$, which is needed in order to have a nontrivial minimum for the potential,
and then $C_1=\lambda^2/(3m^2)$ implies $C_1>0$.

Together, the above conditions give a constraint on the potential ($C_1(\lambda,m^2)$) and a constraint for the 
constant part of the field at infinity to be at the 
minimum of the potential, so that the scalar field solution tends to the nontrivial vacuum,
\be
\frac{1}{2|\Phi|}\frac{dV}{d|\Phi|}(|\Phi_0|)=\frac{dV}{d|\Phi|^2}(|\Phi_0|)=m^2+2\lambda \tilde A^2+3C_1\tilde A^4=0.
\ee
In this case, solving the equations of motion to the next order, to find $\tilde B$, one finally obtains the solution
\be
|\Phi|\sim \sqrt{-\frac{m^2}{\lambda}}+\frac{3}{m^2r^2}+...
\ee
 But also note that in this case, the "Higgs", \emph{i.e.} the fluctuation transverse to the 
vacuum manifold, has zero mass, i.e. the second derivative of the potential at the minimum is also zero,
\be
m_H^2=M^2(|\Phi_0|)\equiv \frac{d^2V}{(d|\Phi|)^2}(|\Phi_0|)=2(m^2+6\lambda |\Phi_0|^2+15 C_1|\Phi_0|^4)= 0.
\ee
We see that in this case we need $\lambda<0, C_1=\lambda^2/(3m^2)>0$. 

It would then seem like we are in the case of type 1 solution, but we are not, since in this case 
\be
V'(|\Phi|)=2|\Phi|\left(m+\frac{\lambda}{m}|\Phi|^2\right)^2\geq 0\;,
\ee
and the potential is strictly increasing, there is no maximum in between $\Phi=0$ and $\Phi_0$ as needed (and $V(0)$ is not larger
than $V(|\Phi_0|)$ as needed).

Thus there is no solution with massless "Higgs" in this case. 

{\em Solution with massive "Higgs"}

But the generic case is one with massive "Higgs". Let us be general, and ask for an arbitrary exponential subleading term (instead of a 
power law) as $r\rightarrow \infty$,
\be
\label{eq:massiveansatz}
|\Phi|\sim \tilde A+B e^{-\a r^\b}.
\ee
Subsituting in the equation of motion, we obtain
\bea
\frac{|\Phi|''}{|\Phi}&=&\frac{B}{\tilde A}e^{-\a r^\b}\left(\a^2 \b^2 r^{2(\b-1)}-\a\b (\b-1) r^{\b -2}\right)+...\cr
=\frac{dV}{d|\Phi|^2}&=&m^2+2\lambda\tilde A^2+3C_1\tilde A^4+4\tilde A Be^{-\a r^\b}(\lambda+3C_1\tilde A^2)+...
\eea
From the vanishing of the constant piece, we again obtain that $\tilde A=|\Phi_0|$, the minimum  (or rather, extremum) of the potential, i.e. 
\be
\frac{dV}{d|\Phi|^2}(|\Phi|=\tilde A=|\Phi_0|)=m^2+2\lambda \tilde A^2 +3C_1\tilde A^4=0\;,\label{vacuum}
\ee
as in the previous case. 

From the exponential term, imposing that we are {\em not} in the massless Higgs case, i.e. we have 
\be
\lambda+3C_1\tilde A^2\neq 0\;,
\ee
which amounts to the condition on the coefficients of the potential
\be
C_1\neq -\frac{\lambda}{3\tilde A^2}\Rightarrow C_1 \neq \frac{\lambda^2}{3m^2}\;,
\ee
then we need $\b=1$ (since $2\b -2>\b -2$ if $\b>0$, so the first term on the left hand side of the equation of motion is dominant, and to 
match the right hand side, it must be a constant of $r$, i.e. $\b=1$). 

Then, equating the coefficients of the exponential terms in the equations of motion, we obtain
\bea
\frac{\a^2 B}{\tilde A}&=&4\tilde A B (\lambda+3C_1\tilde A^2)\Rightarrow \cr
\a^2&=& 4(\lambda \tilde A^2+3C_1 \tilde A^4)=4(-\lambda \tilde A^2 -m^2)\;,\label{alpha}
\eea
where in the last equality we have used the equation for $\tilde A$ to be a minimum of the potential, (\ref{vacuum}). 

If $C_1=0$, from (\ref{vacuum}) we get $\lambda\tilde A^2=-m^2/2$, which when substituted in $\a^2$ above leads to $\a^2<0$, which is 
impossible, since $|\Phi|$ is real and positive. 

Therefore we need $C_1\neq 0$, in which case the solution of the extremum condition (\ref{vacuum}) is 
\be
|\Phi_0|^2=\tilde A^2 =\frac{-\lambda\pm \sqrt{\lambda^2-3C_1m^2}}{3C_1}.\label{tildeA}
\ee
Note that we expect the smallest solution (with the minus sign) to be a maximum, and then the largest solution (with the plus sign) to be the 
minimum. That is so, since $|\Phi|$ starts to increase from zero, so needs to encounter a maximum before a minimum. 

From the condition that the above is real and positive (such that $|\Phi|$ is real and positive) we get first $\lambda^2-3C_1m^2>0$, i.e.
\be
C_1<\frac{\lambda^2}{3m^2}.
\ee
Note that in this inequality we have excluded the case $C_1=\lambda^2/3m^2$, since that was the massless Higgs case above. 

Substituting $\tilde A^2$ from (\ref{tildeA}) in $\a^2$ in (\ref{alpha}), we get
\be
\a^2=4\frac{(\lambda^2-3m^2 C_1)\mp \lambda\sqrt{\lambda^2-3C_1m^2}}{3C_1}.\label{alph}
\ee
We note that this is real under the same condition as $\tilde A^2$. 

We still need to impose that $\tilde A^2>0$ and $\a^2>0$. 

{\bf a) If $C_1>0$,} $\tilde A^2>0$ implies $\lambda<0$, in which case both solutions are OK for $\tilde A^2$, but we easily see that then only the
upper one is OK for $\a^2>0$. That is as it should be, since the solution with the upper sign corresponds to the minimum, where the field at 
infinity should end up, whereas the lower sign corresponds to the maximum, which shouldn't be allowed as solution at infinity. 

\begin{figure}[ht!]
\centering{
\subfigure[]{
\includegraphics[width=.45\textwidth]{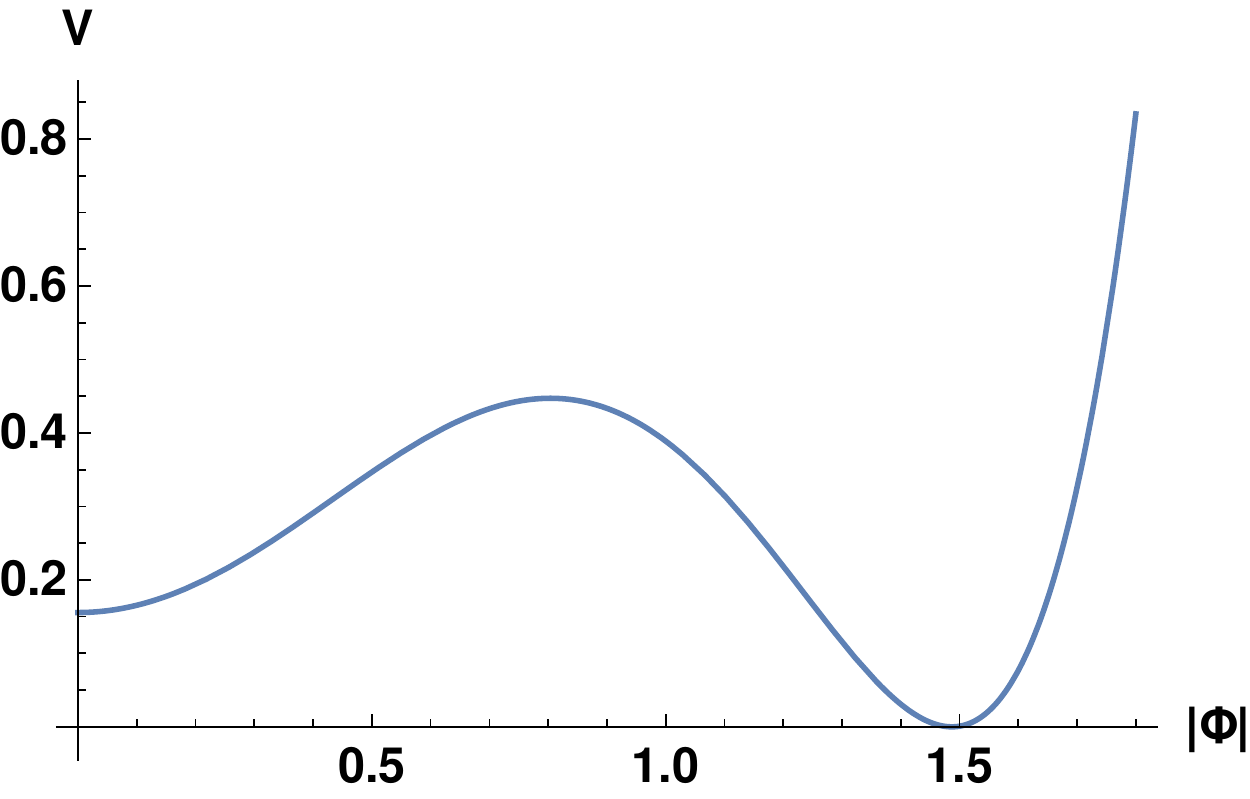}
\hspace{0.5cm}}       
\subfigure[]{
\includegraphics[width=.45\textwidth]{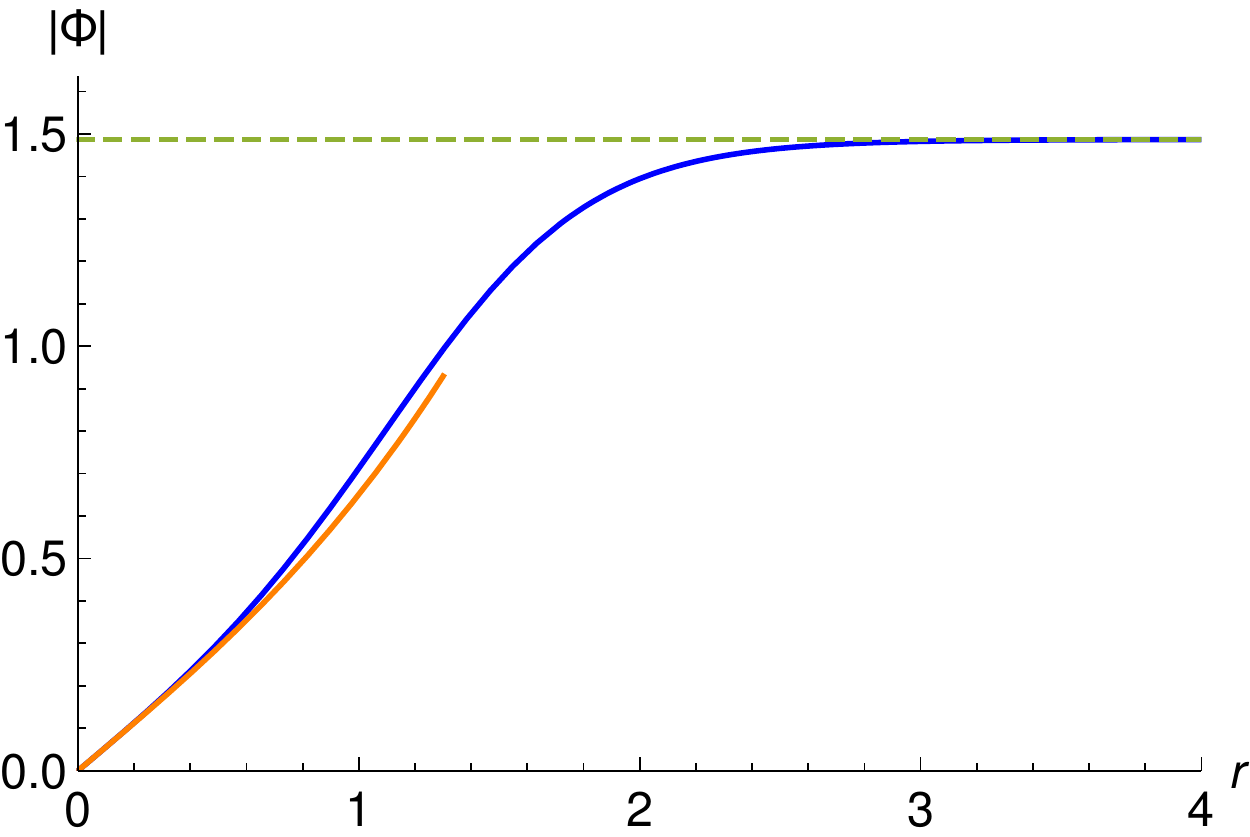}}
}
\caption{\small In (a) a type-2 potential admitting vortex solutions is shown. Depicted in (b) is the vortex profile (in blue) obtained by numerically solving the field equations for $|\Phi(r)|$. As expected, $|\Phi(r)|$ asymptotically approaches the global minimum of the potential $|\Phi_0| \sim 1.49$ for large values of $r$. Here we have also included (in orange) the analytic solution $|\Phi(r)|\sim Ar(1+m^2r^2/6)$ obtained in \eqref{phirzero} valid for $r\sim 0$, which also shows its improving accuracy the closer we get to $r=0$. 
\label{fig:type2vortex}}
\end{figure}

Thus we have 
\be
\lambda<0;\;\;\; 0<C_1<\frac{\lambda^2}{3m^2}.\label{lambdaC1}
\ee
This potential and a vortex solution obtained by numerically solving the equations of motion are shown in Fig.~\ref{fig:type2vortex}.

{\bf b) If $C_1<0$,} $\tilde A^2>0$ implies $\lambda>0$, in which case only the lower sign is OK for $\tilde A^2>0$, but then only the upper one is 
OK for $\a^2>0$, so we have no good solution. 

Since we are in the case of type 1 solution, where in the inverted potential problem one starts off with kinetic energy, and in the end it is all converted to 
potential energy, it means that we need to have $V(|\Phi_0|)<V(0)$, which leads to the condition 
\be
\frac{|\Phi_0|^2}{3}(\lambda|\Phi_0|^2+2m^2)<0\Rightarrow \frac{-\lambda^2+\lambda\sqrt{\lambda^2-3C_1m^2}+6m^2 C_1}{3C_1}<0\;,
\ee
or in the end
\be
0<C_1<\frac{\lambda^2}{4m^2}.\label{lambdaC1p}
\ee

Finally, we need to impose that the asymptotic value of $|\Phi|$ is a minimum, not a maximum of the potential. For that, we need to impose that
the mass squared, the second derivative of the potential at the mininum, is positive. We have
\be
M^2\equiv \frac{d^2V}{(d|\Phi|)^2}=2(m^2+6\lambda |\Phi|^2+15 C_1|\Phi|^4)\geq 0.\label{masssq}
\ee

But note that, using the equation for the minimum (\ref{vacuum}), 
\be
M^2=2(4\lambda\tilde A^2+12C_1\tilde A^4)=2\a^2\;,
\ee
so we have already imposed its positivity, and moreover, since $M=m_H$ is the mass of the "Higgs", i.e. of the physical excitation transverse
to the vacuum manifold, and denoting $\tilde A\equiv \Phi_0$, we have
\be
|\Phi|\sim |\Phi_0|+B e^{-\frac{m_H r}{\sqrt{2}}}+...\label{behinf}
\ee
Thus as usual for vortices, the decay of the scalar field at infinity is governed by the mass of the "Higgs".
Also as usual, $B$ is only fixed by the "shooting method", which means we vary $B$ (defining the asymptotics at infinity) 
until the solution has the right asymptotics at $r=0$. 

In conclusion, in this case we need the conditions (\ref{lambdaC1}), (\ref{lambdaC1p}) on the parameters of the potential in order to have a vortex solution.

We have already seen that we need $C_1\neq 0$ if $m^2\neq 0,\lambda\neq 0$, and it is easy to see that if $C_1=0$ and 
$m^2=0$ or $\lambda=0$ (purely quartic or purely quadratic potential) we also cannot have a nontrivial soliton solution, since we are in the 
case of type 1 or 2 (or type 3 or  4 for the compacton), which do not happen for purely quartic or purely quadratic potentials. 
Therefore in the following we will assume $C_1\neq 0$, and consider separetely the cases $\lambda=0$ and $m^2=0$.

{\bf B. \underline{$C_1\neq 0, m^2=0, \lambda\neq 0$}}. 

In this case, 
\be
V=C_1|\Phi|^6+\lambda|\Phi|^4.
\ee

As seen in \cite{Murugan:2014sfa}, considering the same vortex asymptotics near $r\rightarrow 0$ as before, 
\be
|\Phi|\sim Ar+Cr^p+...\label{rzero}
\ee
from the equations of motion we obtain 
$p=5$ and $C=\lambda \frac{A^3}{10}$ so that as $r\rightarrow 0$,
\be
|\Phi|\sim Ar\left(1+\frac{\lambda A^2}{10}r^4+...\right)\, .\label{phizero}
\ee

For the behaviour at $r\rightarrow\infty$, we can actually take a $m^2\rightarrow 0$ limit on the above case (with $m^2> 0$). 

We first see that the power law behaviour at 
infinity (massless Higgs) is impossible, since $C_1=\lambda^2/(3m^2)$ doesn't have a good $m^2\rightarrow 0$ limit. We can also check 
directly that the power law ansatz doesn't solve the equations of motion. 

Then we can take the $m^2\rightarrow 0$ of the massive Higgs case. From the condition (\ref{vacuum}), we have a nontrivial vacuum at
\be
|\Phi|=|\Phi_0|=\tilde A=\sqrt{-\frac{2\lambda}{3C_1}}.
\ee
We also have a trivial vacuum at $|\Phi|=0$ (see also (\ref{tildeA}) for $m=0$), which is now a {\em maximum } of the potential.
Thus we are now in the case of type 2 vortex in our classification.

The mass of the Higgs (around the nontrivial vacuum) is, from (\ref{masssq}),
\be
m_H^2\equiv M^2=\frac{d^2V}{(d|\Phi|)^2}(|\Phi_0|)=\frac{16\lambda_1}{3C_1}.
\ee
We need this to be positive, so that $|\Phi_0|$ is a minimum, not a maximum. This again imposes $C_1>0$, which in turn, from the form of 
$|\Phi_0|$, implies that $\lambda<0$. 

Then from (\ref{alph}) again we have for the nontrivial exponent,
\be
\a^2=\frac{m_H^2}{2}=\frac{8\lambda^2}{2C_1}\;,
\ee
as well as the trivial $\a=0$. 

So the nontrivial behaviour at infinity is again given by the general form (\ref{behinf}), and the coefficient $B$ would be fixed by the shooting method, 
by imposing that we have the correct behaviour (\ref{phizero}) at $r\rightarrow 0$. 

Note that we {\em cannot} impose that $|\Phi|$ goes over to the trivial vacuum $|\Phi|=0$ at $r\rightarrow \infty$. This was what 
was assumed in \cite{Murugan:2014sfa}, but it is incorrect, since this would not correspond to one of the cases (1, 2, 4 or 5) that 
we have described.

{\bf C. \underline{$C_1\neq 0, \lambda=0, m^2\neq 0$}}

The behaviour at $r\rightarrow 0$ is again given by (\ref{phirzero}), since that was independent of $\lambda$. 

Now again there is no solution with massless Higgs, since $|\Phi_0|^2=-m^2/\lambda$ doesn't have a $\lambda\rightarrow 0$ limit. 

From the condition (\ref{vacuum}), we have an extremum at 
\be
|\Phi|^2=|\Phi_0|^2=\tilde A^2=\sqrt{-\frac{m^2}{3C_1}}\;,
\ee
which requires $C_1<0$ (since $m^2>0$ now), but then from (\ref{masssq}), 
\be
M^2(|\Phi_0|)=\frac{d^2V}{(d|\Phi|)^2}(|\Phi_0|)=-8m^2\sqrt{-\frac{m^2}{3C_1}}<0\;,
\ee
so the extremum is actually a maximum. Indeed, in this case, the potential grows to a maximum and then drops without bound.
Since there is no minimum of the potential, there is no nontrivial vacuum.
In fact, since $C_1<0$, $V(|\Phi|\rightarrow \infty)=-\infty$, so the potential is unbounded from below. 

Then at $r\rightarrow \infty$, we could only imagine that the field tends to the trivial vacuum, however 
then the equation of motion gives a contradiction, 
\be
\frac{n(n+1)}{r^2}=m^2\;,
\ee
which is consistent with the fact that there is no vortex case corresponding to this possibility, according to our general analysis. 

{\bf D. \underline{$C_1\neq 0, m^2=\lambda=0$}}

This is a purely sextic potential, $V(|\Phi|)=C_1|\Phi|^6$. This choice was analyzed in \cite{Murugan:2014sfa}. 
On physical grounds we must choose $C_1>0$, otherwise the potential is negative definite, and unbounded from below. 
There is no usual vortex case corresponding to this potential. Nevertheless, we consider a more unusual possibility, since in this 
case we can solve the equations of motion exactly.

The possible behaviour at $r\rightarrow 0$ consistent with a normal vortex ansatz is easily found to be 
\be
|\Phi|\sim Ar\left(1+\frac{C_1A^4}{14}r^6+...\right)\;,\label{rtoinf}
\ee
whereas at $r\rightarrow \infty$ we must have 
\be
|\Phi|\sim \frac{1}{\left(4C_1\right)^{1/4}\sqrt{r}}\, .
\ee

In this case we can actually solve exactly the equation of motion. It integrates to 
\be
r+K_2/\sqrt{C_1}=\pm \int\frac{d|\Phi|}{\sqrt{C_1|\Phi|^6+K_1}}\;,
\ee
but as explained in \cite{Murugan:2014sfa}, for $C_1<0$, the potential is unbounded from below, and for the physical case $C_1>0$, there is no 
solution with the normal vortex asymptotics at zero, i.e. no solution with $|\Phi|(r=0)=0$. Instead, there is a kind of vortex solution
with $|\Phi|(r=0)\neq 0$, obtained by putting $K_1=0$ in the above, namely
\be
|\Phi|=\frac{1}{\sqrt{\sqrt{C_1}2r+2K_2}}\, ,\label{phiatzero}
\ee
We see that it has the the right $r\rightarrow \infty$ asymptotics (\ref{rtoinf}), but it has $|\Phi|(0)=1/\sqrt{2K_2}$, and 
finite derivative at zero,
\be
|\Phi|'(0)=-\frac{\sqrt{C_1}}{\sqrt{2K_2}}\, .
\ee
However, this probably doesn't make sense for our vortex equations, since the rest of the equations are satisfied at $r=0$, but 
$|\Phi|(0)\neq 0$ contradicts the equations of motion at $r=0$. 

{\bf II. \underline{$m^2<0$}}

In this case, if $m^2\neq 0$ (the case $m^2=0$ was already analyzed at type $I$ $B$ above), 
the only possibilities are vortices of type 2 or 3. Since we will not consider compactons until the end of the subsection, 
we consider type 2 only.

{\bf A. \underline{$C_1\neq 0, \lambda\neq 0, m^2\neq 0$}. }

Again, near the $r\sim 0$ region we have
\be
\label{phirzeronegm2}
|\Phi| \sim A r\left(1 - \frac{|m^2|r^2}{6} + \cdots \right)
\ee
Just as in the $m^2>0$ case, we will now analyze the $r \to \infty$ behaviour. 

{\em Solution with massless "Higgs"}

As before, assuming a non-trivial vacuum with a power series decay for large $r$ of the form
\be
|\Phi| \sim \tilde A + \frac{\tilde B}{r} + \cdots
\ee 
When plugging this into the equations of motion we obtain the same conditions as in the $m^2>0$ case, namely,
\be
\label{negm2_cond}
m^2 + 2 \lambda \tilde A^2 + 3C_1\tilde A^4=0 \quad \text{and} \quad \lambda = -3C_1 \tilde A^2
\ee
which when combined yield
\be
\lambda^2 = 3C_1 m^2.
\ee
 However, we can see that there is a substantial difference. Since $m^2$ is negative, so must $C_1$ be, otherwise $\lambda $ becomes imaginary. Thus, we need $C_1=  \lambda^2/(3m^2) <0$, and from \eqref{negm2_cond} we also need $\lambda >0$. Thus, the potential is unbounded from below. Also, once again, the second derivative of the potential at $|\Phi| =  \tilde A$, \emph{i.e}, the mass of the putative ``Higgs'' is exactly zero. 

Unlike the $m^2>0$ case however, this solution is actually possible (if we ignore the fact that the potential is unbounded from below, hence unphysical). 
Indeed, now 
\be
V'(|\Phi|)=2|\Phi|m^2\left(1+\frac{\lambda}{m^2}|\Phi|^2\right)\leq 0\;,
\ee
so the potential is monotonically decreasing, and has $V'=0$ at $|\Phi_0|=\sqrt{-m^2/\lambda}$, which is therefore a good point for 
the $r\rightarrow \infty$ behaviour. This satisfies all the conditions on the type 2 vortex solution. 

{\em Solution with massive "Higgs"}

The ansatz is given in \eqref{eq:massiveansatz} which we repeat here for convenience
\be
|\Phi|\sim \tilde A+B e^{-\a r^\b}.
\ee
Just as before, plugging this into the equation of motion for $|\Phi|$ gives $\beta=1$ plus the conditions
\be
m^2 + 2 \lambda \tilde A^2 + 3 C_1 \tilde A^4=0\,, \quad \text{and} \quad \alpha^2 =4 \tilde A^2(\lambda + 3C_1 \tilde A^2)
\ee
at leading and subleading order respectively. Combining both yields
\be
\alpha^2 = -4(m^2+\lambda \tilde A^2)
\ee
Since here $m^2<0$, demanding positivity of $\alpha^2$ gives $\lambda < |m^2|/\tilde A^2$. Note that contrary to the $m^2>0$ case, 
$C_1=0$ is not ruled out here. More precisely, combining both leading and subleading conditions with $C_1=0$ yields $\alpha^2=-2m^2$, 
and since $m^2<0$, the positivity of $\alpha^2$ is ensured.

Solving for $\tilde A^2$ yields
\be
\label{eq:A2negm2}
|\Phi_0|^2 = \tilde A^2 = \frac{-\lambda \pm \sqrt{\lambda^2 + 3 |m^2|C_1}}{3C_1}\;,
\ee
and we must impose that  $\tilde A^2$ be positive. 

\begin{figure}
  \centering
   \includegraphics[scale=0.3]{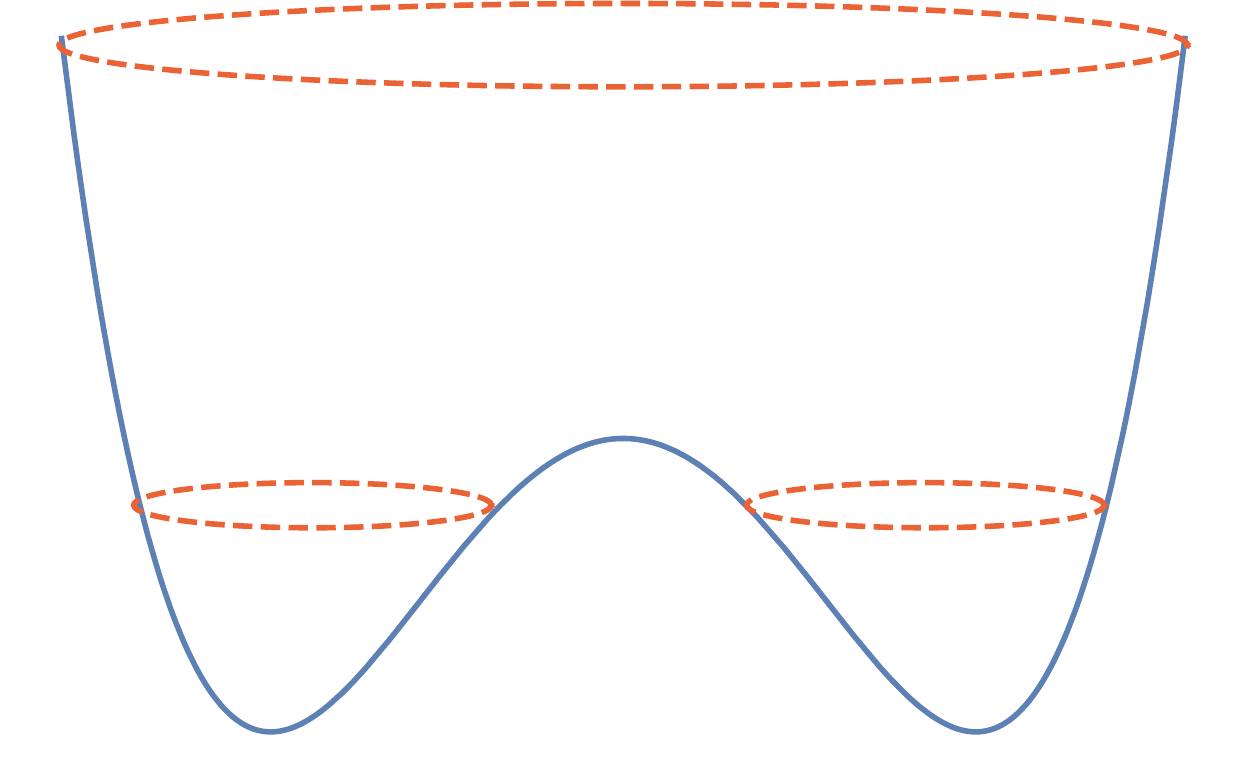}
   \caption{The potential for the $m^2<0$, $\lambda=0$ or $\lambda\neq 0$, $C_1>0$ case.}
   \label{fig:two}
\end{figure}

{\bf a) If $C_1>0$,} then the requirement $\tilde A^2>0$ imposes no conditions on $\lambda$. Although it may look like there is upper bound on 
$\lambda$ coming from the $\a^2>0$  condition $\lambda < |m^2|/\tilde A^2$, when solving for $\tilde A^2$ one can easily see that $\lambda$ 
can take any value, since we obtain 
\be
\a^2=\frac{4\sqrt{\lambda^2+3C_1|m^2|}}{3C_1}(\sqrt{\lambda^2+3C_1|m^2|}\mp \lambda)>0.
\ee
This is in contrast with the $m^2>0$ case where $\lambda$ had to be negative. Also here only the upper sign in \eqref{eq:A2negm2} is OK, and
the condition $\alpha^2>0$ is also always satisfied. Thus we have
\be
C_1>0, \, \lambda \in \mathbb{R}.
\ee
In this case $|\Phi_0|$ is a minimum, and we are in the case of type 2 vortex. The potential is sketched in Fig. \ref{fig:two}.

\begin{figure}[t]
  \centering
   \includegraphics[scale=0.3, angle=180]{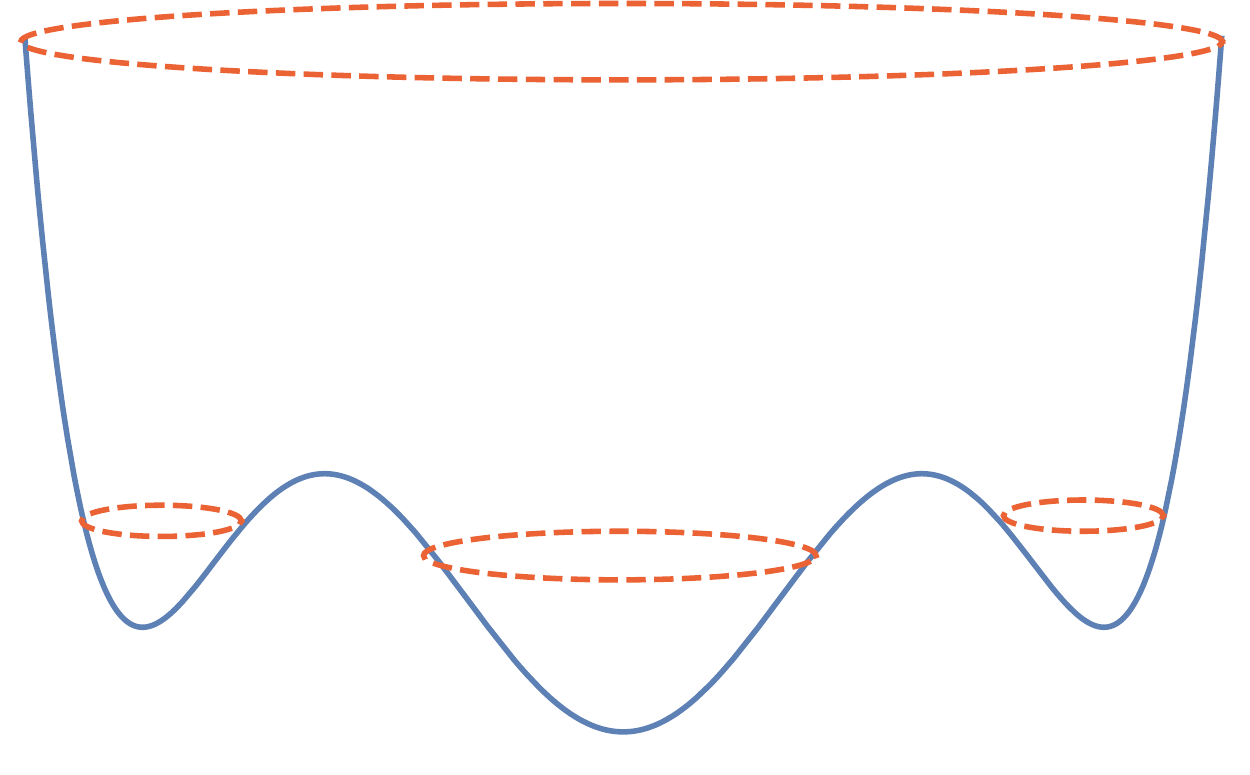}
   \caption{The potential for the $m^2<0$, $C_1<0$ case.}
   \label{fig:inv-three}
\end{figure}

{\bf b) If $C_1<0$,} the reality condition for $\tilde A^2$ imposes that $\lambda^2 > 3|m^2 C_1|$. This is also the same condition for which 
$\alpha^2>0$, but now only the lower sign is OK (instead of only the upper sign for $C_1>0$). Note that the equality 
$\lambda^2 = 3|m^2 C_1|$ has been left out because it 
corresponds to the massless ``Higgs'' case 
previously analyzed.  Demanding $\tilde A^2>0$ imposes that $\lambda <0$. Although $C_1<0$ implies that the potential is unbounded 
from below, there is a (unique) local minimum at
\be
|\Phi_0| =  \tilde A^2= \frac{|\lambda|-\sqrt{\lambda^2-3|m^2 C_1|}}{3|C_1|}.
\ee
The potential is shown in Figure~\ref{fig:inv-three}.
Thus, we need
\be
C_1<0;\quad 0<\lambda  < \sqrt{3|m^2 C_1|}\;,
\ee
and we are still in the case of vortex of type 2. 

{\bf B. \underline{$C_1\neq 0, \lambda= 0, m^2\neq 0$}. }

Here
\be
V=m^2 |\Phi|^2+ C_1 |\Phi|^6
\ee
The behaviour near $r=0$ is the one given in \eqref{phirzeronegm2}, and once again there's no solution with a massless ``Higgs''. Using \eqref{eq:massiveansatz} for the behavior for large $r$, we have
\be
|\Phi_0|=\tilde A^2 = \sqrt{\frac{|m^2|}{C_1}}
\ee
from where we immediately see that $C_1>0$. In that case, we see that in fact this is simply a subset of point A a) above, which had $\lambda\in \mathbb{R}$, 
which includes zero.
The subleading term in the large $r$ expansion of the equation of motion for $|\Phi|$ gives 
\be
\alpha^2 = -4m^2,
\ee
thus, $\alpha^2>0$ is automatically satisfied. We can also see that the extremum of the potential at $|\Phi|=|\Phi_0|$ is also a minimum:
\be
M^2 (|\Phi_0|)= \frac{d^2 V}{(d|\Phi|)^2}(|\Phi_0|) = -4m^2>0
\ee
The shape of the potential is shown in Figure~\ref{fig:two}.
Thus,  we only need
\be
C_1 >0.
\ee

{\bf C. \underline{$C_1=0, \lambda \neq 0, m^2\neq 0$}. }

We mentioned earlier that, contrary to the $m^2>0$ case, $m^2<0$ allows $C_1=0$. The potential is
\be
V=m^2 |\Phi|^2+ \lambda |\Phi|^4
\ee
The behaviour near $r=0$ is again given by \eqref{phirzeronegm2}. Imposing the behaviour $|\Phi|\sim \tilde A+\frac{\tilde B}{r^n}+...$ is again inconsistent. Thus we try
\be
|\Phi|\sim \tilde A+B e^{-\a r^\b}.
\ee
For this ansatz everything works out, which is of course not surprising since it's the usual Higgs mechanism. At leading order in the large $r$ expansion we obtain
\be
m^2 + 2 \lambda \tilde A^2=0 \quad \Rightarrow \quad \tilde A^2 = -\frac{m^2}{2\lambda}
\ee
which requires $\lambda >0$, yielding a bounded-from-bellow potential. At subleading order we obtain once again $\beta=1$ and 
\be
\alpha^2 = -2m^2
\ee
thus, the positivity of $\alpha^2$ is also immediate. The Higgs mass is again $m_H=2|m|$, thus the solution is 
\be
|\Phi|\sim |\Phi_0|+B e^{-\frac{m_H r}{\sqrt{2}}}+...\label{behinf}
\ee
In conclusion, we only need
\be
\lambda >0\;,
\ee
and we are again in the case of type 2 vortex.

{\bf Compacton solutions}

Finally, we consider compacton solutions, which are easiest to obtain. In fact, the possibilities of type 3 or 4 cover most potentials. 

To have a compacton solution of type 3, we only need to have the leading term at $|\Phi|\rightarrow 0$ in the potential be negative. 
If $m^2\neq 0$, the condition is simply $m^2<0$, and then we will always have a compacton solution, with a sufficiently small initial velocity, i.e. a 
sufficiently small $A$ in $|\Phi|\sim A r$ as $r\rightarrow 0$. If $m^2=0$, then the condition is $\lambda<0$, and if also $\lambda=0$, then $C_1<0$. 

To have a compacton solution of type 4, we need that the leading term is positive, and the subleading is negative (and the final subleading term is not 
positive and large, so as to reverse it). So if $m^2\neq 0$, the condition 
is that $m^2>0$, but $\lambda<0$ and $C_1<0$ or $C_1>0$ but very small, or that $\lambda=0$ and $C_1<0$. If $m^2=0$, the condition is that $\lambda>0$
and $C_1<0$. In fact, the only condition in this case is that if $m^2>0$, then $V(|\Phi_0|)=0$ has also a nontrivial solution, besides the trivial one 
$|\Phi_0|=0$, so, since the equation has the solution
\be
|\Phi_0|=\frac{-\lambda\pm \sqrt{\lambda^2-4C_1m^2}}{2}\;,
\ee
we need that $C_1<0$, or $\lambda\leq 0$ and $0<C_1<\lambda^2/m^2$. 

\begin{figure}
\centering{ 
\subfigure{
\includegraphics[width=.4\textwidth]{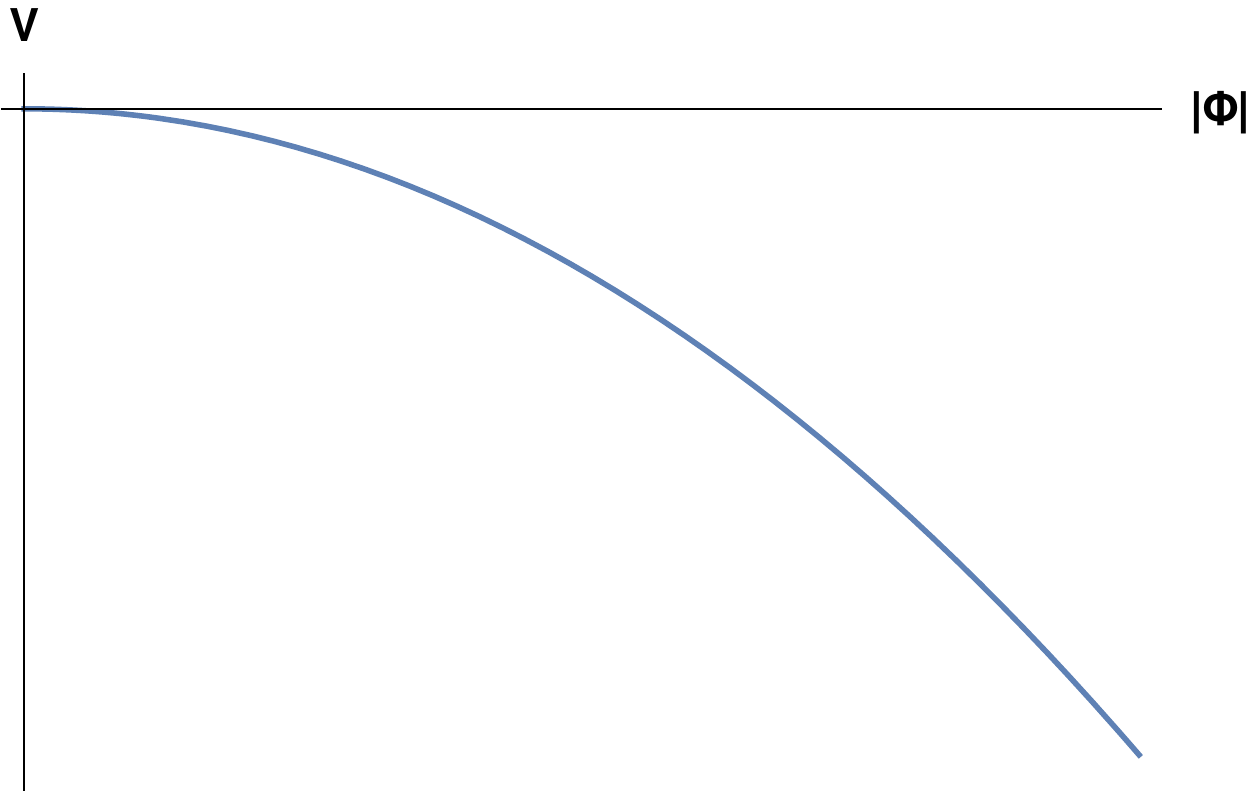}
\hspace{1cm}}
\subfigure{
\includegraphics[width=.4\textwidth]{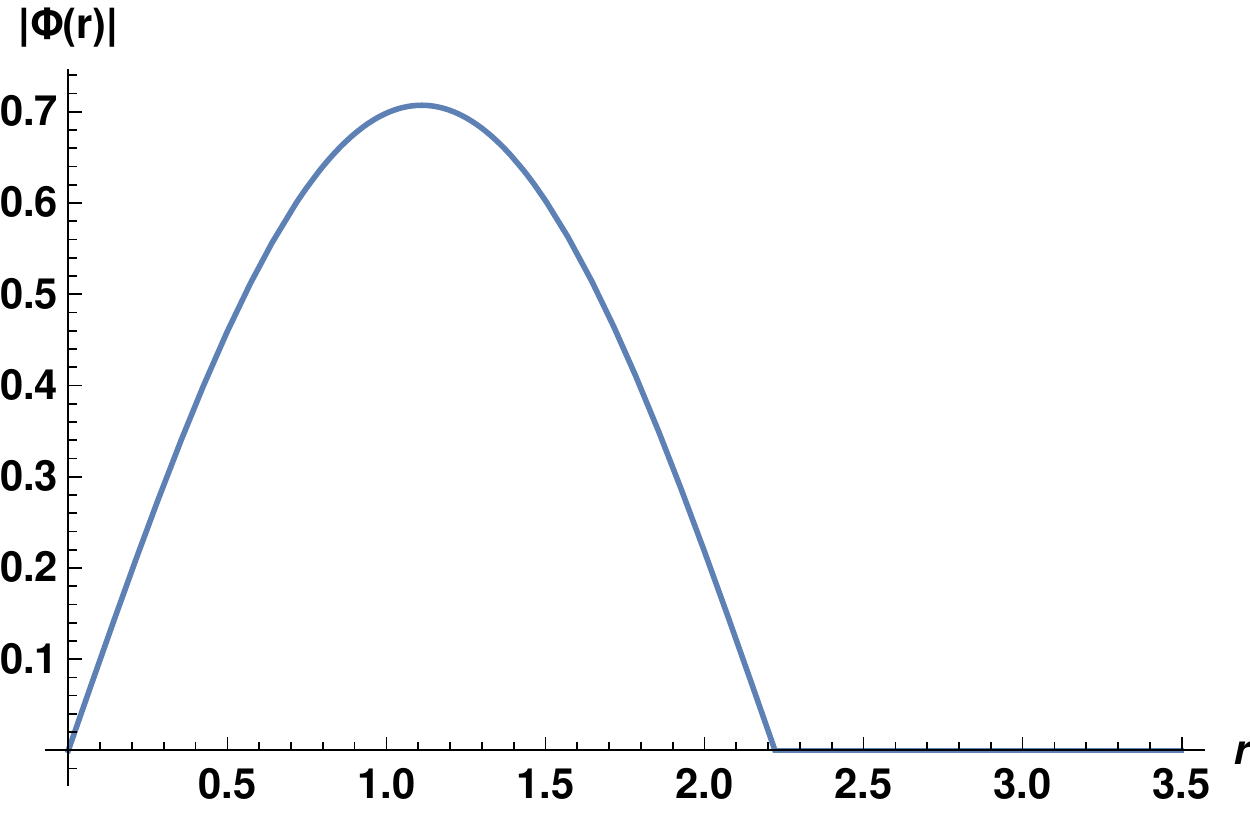}}
}\\
\centering{ 
\subfigure{
\includegraphics[width=.4\textwidth]{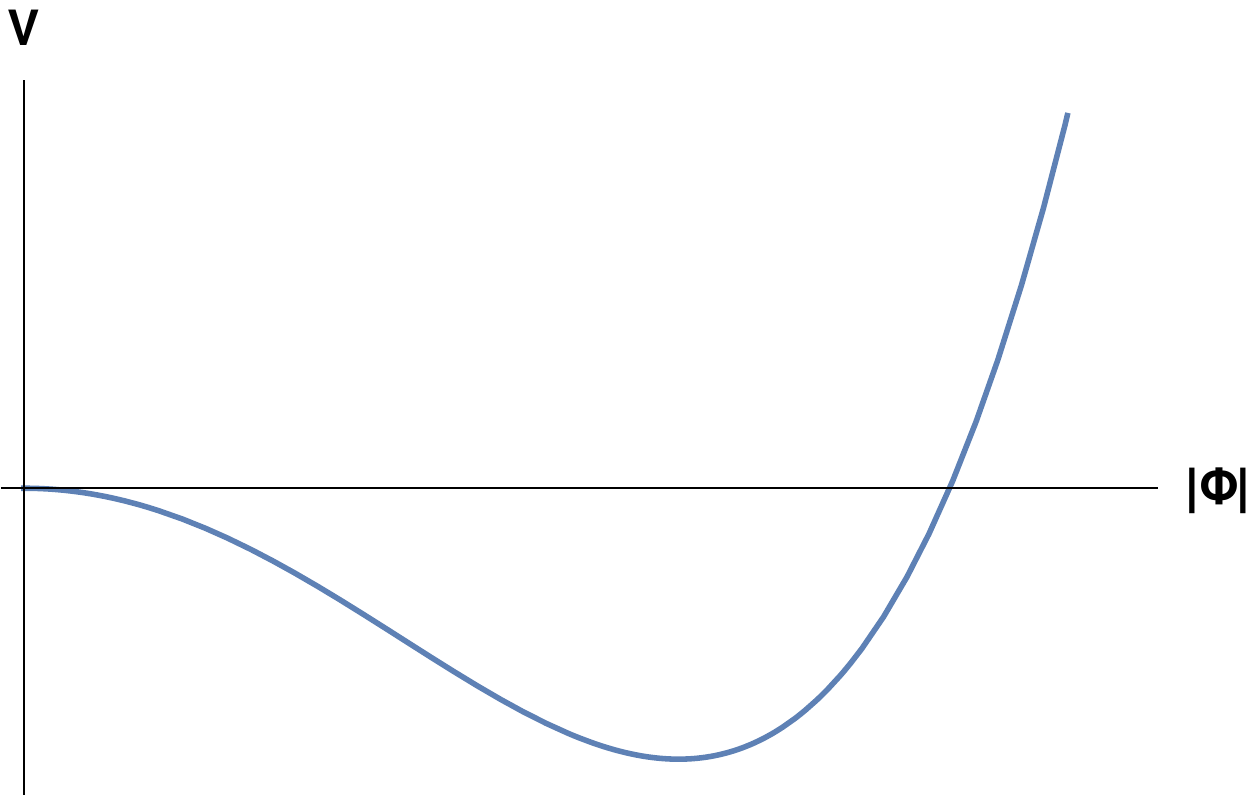}
\hspace{1cm}}
\subfigure{
\includegraphics[width=.4\textwidth]{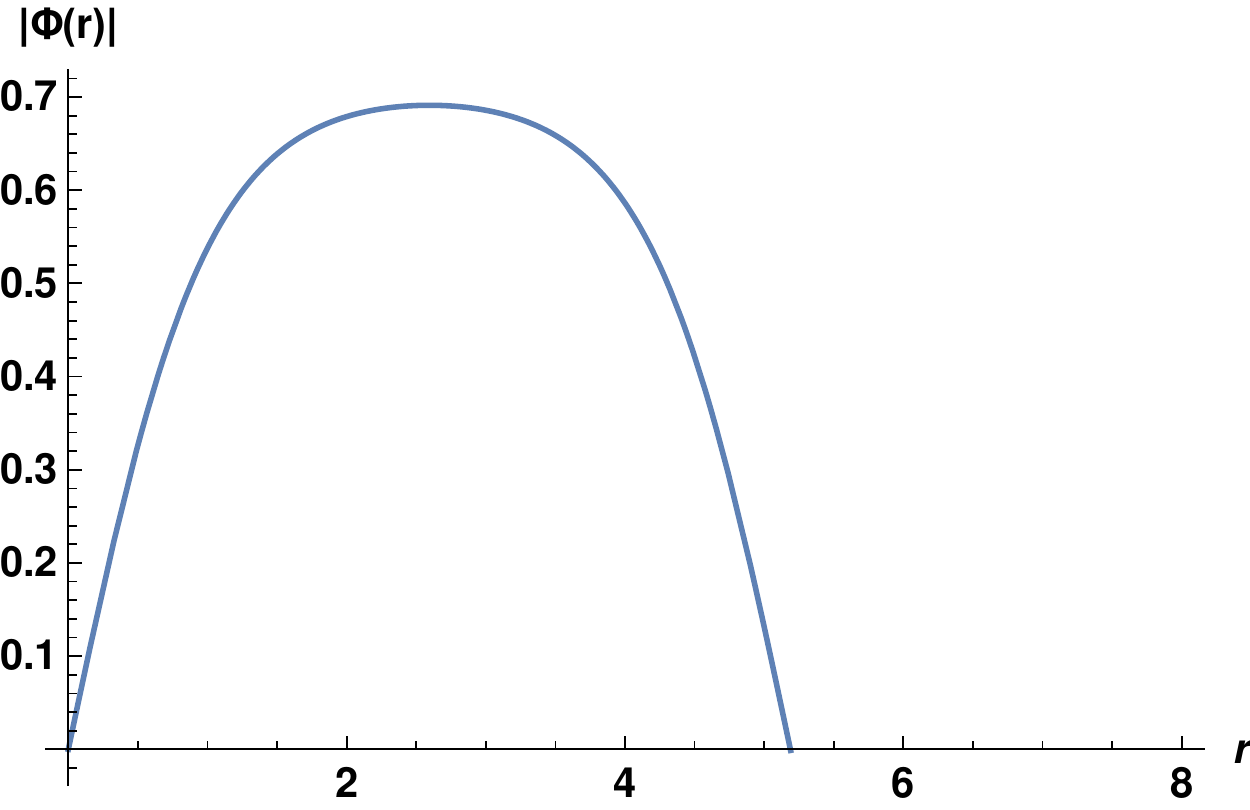}}
}\\
\centering{ 
\subfigure{
\includegraphics[width=.4\textwidth]{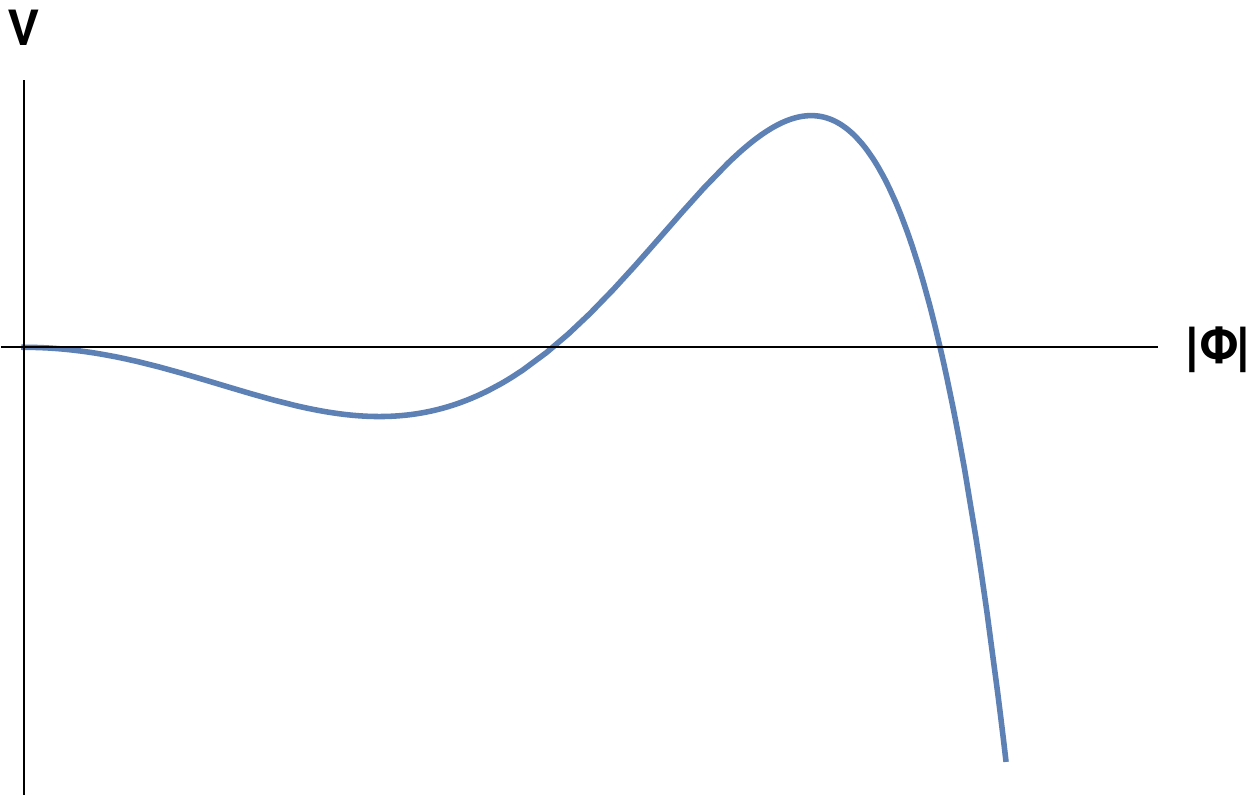}
\hspace{1cm}}
\subfigure{
\includegraphics[width=.4\textwidth]{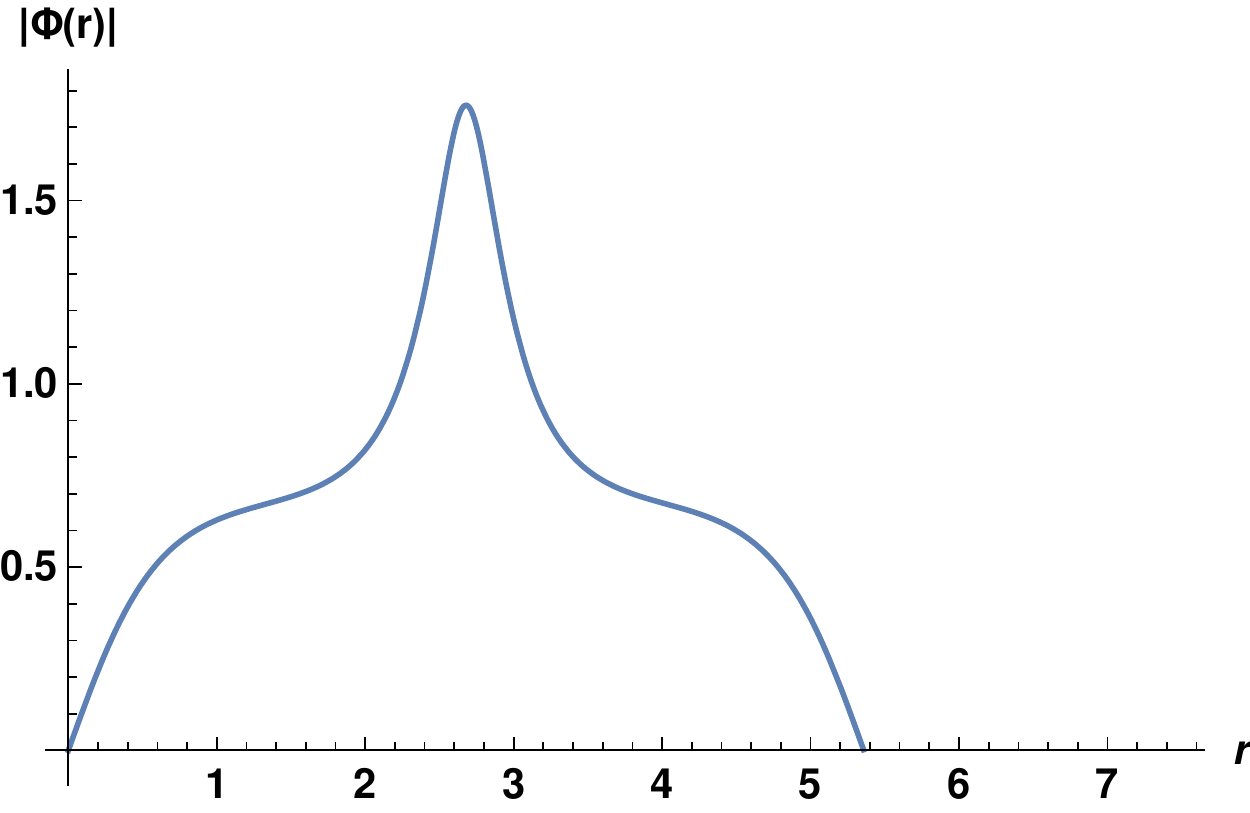}}
}\\
\centering{ 
\subfigure{
\includegraphics[width=.4\textwidth]{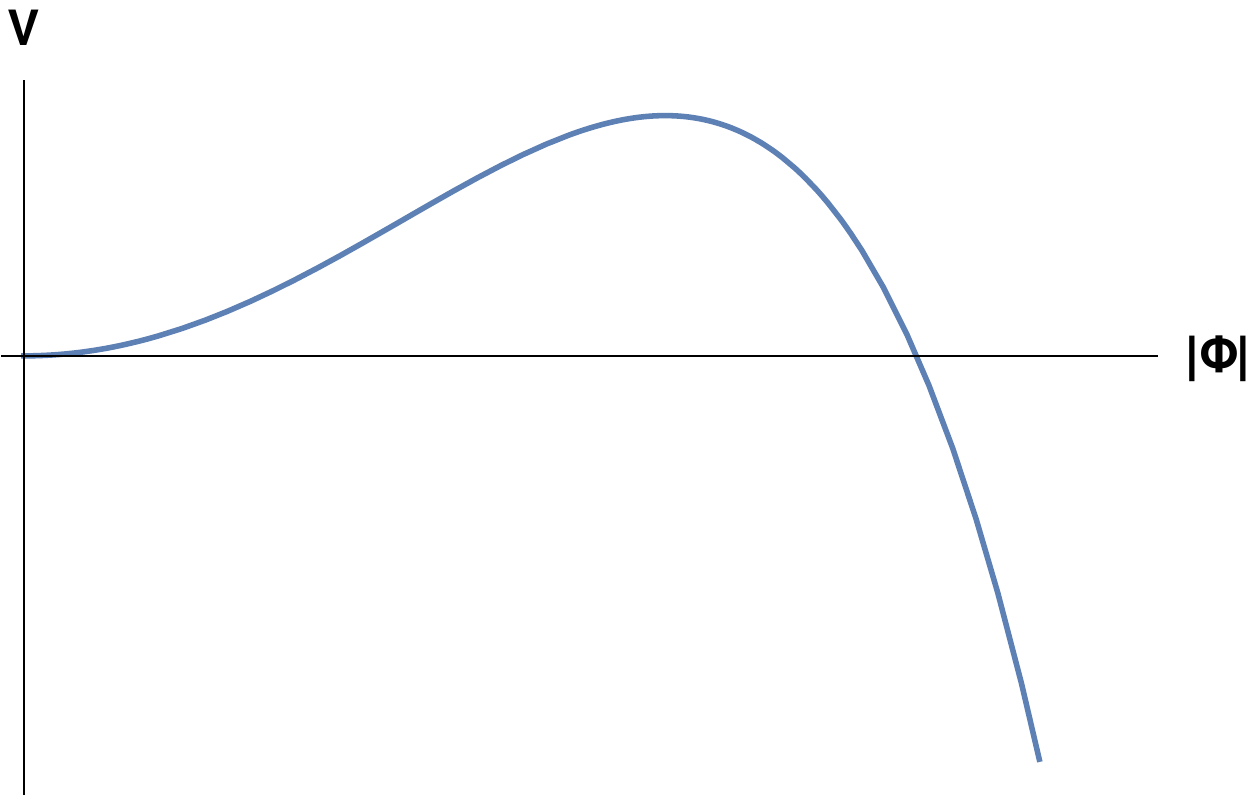}
\hspace{1cm}}
\subfigure{
\includegraphics[width=.4\textwidth]{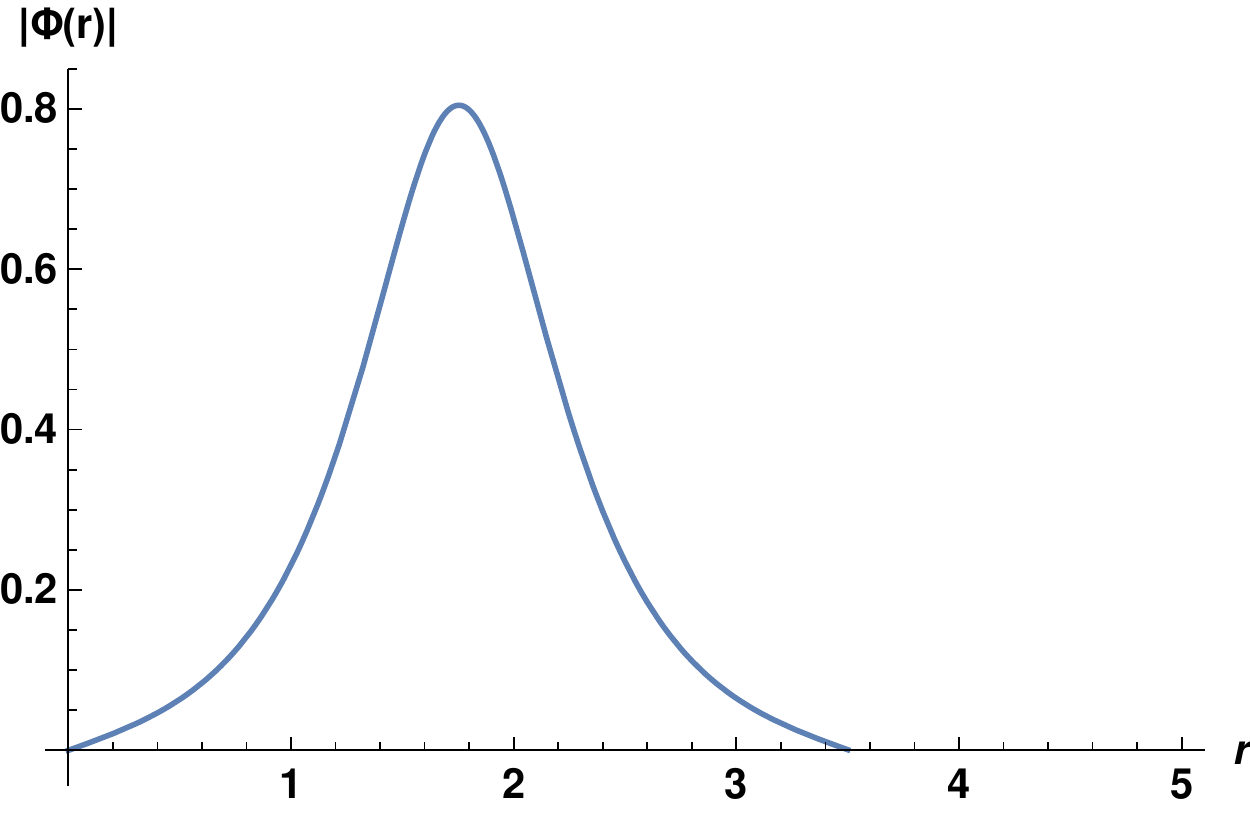}}
}\\
\centering{ 
\subfigure{
\includegraphics[width=.4\textwidth]{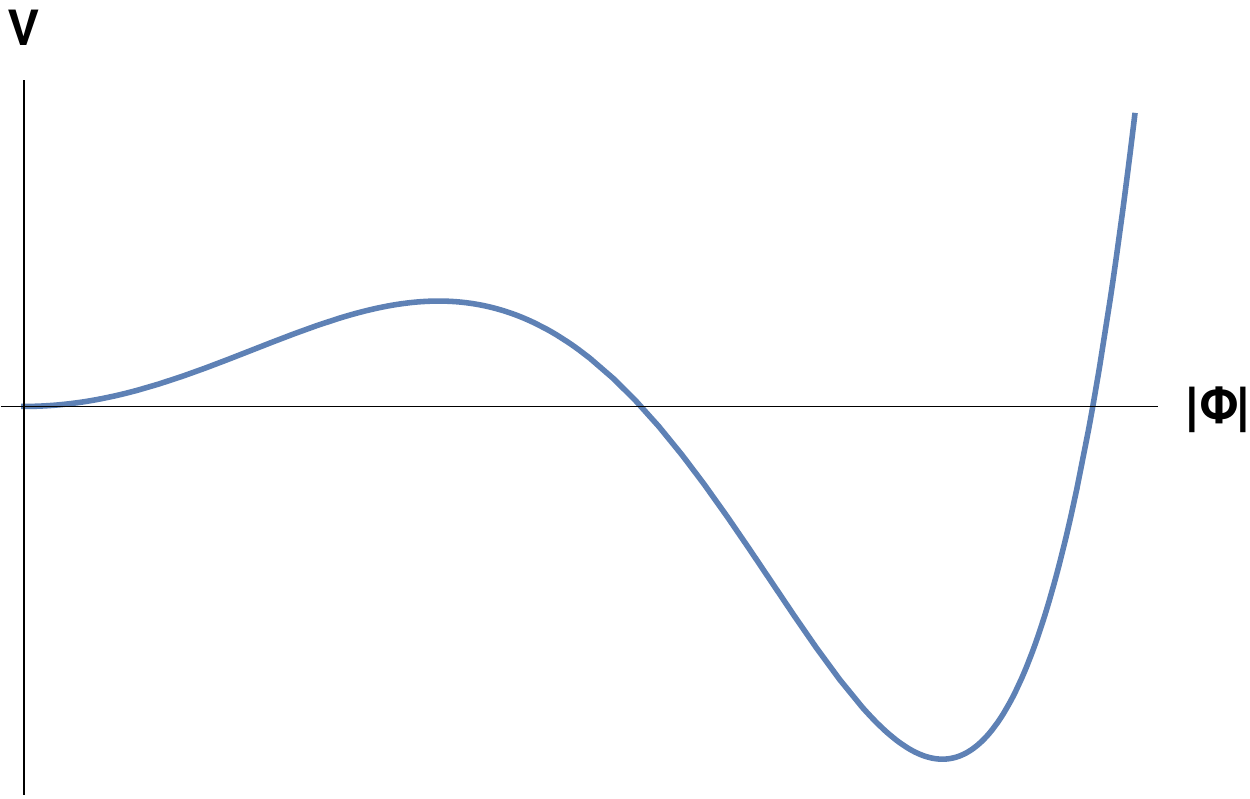}
\hspace{1cm}}
\subfigure{
\includegraphics[width=.4\textwidth]{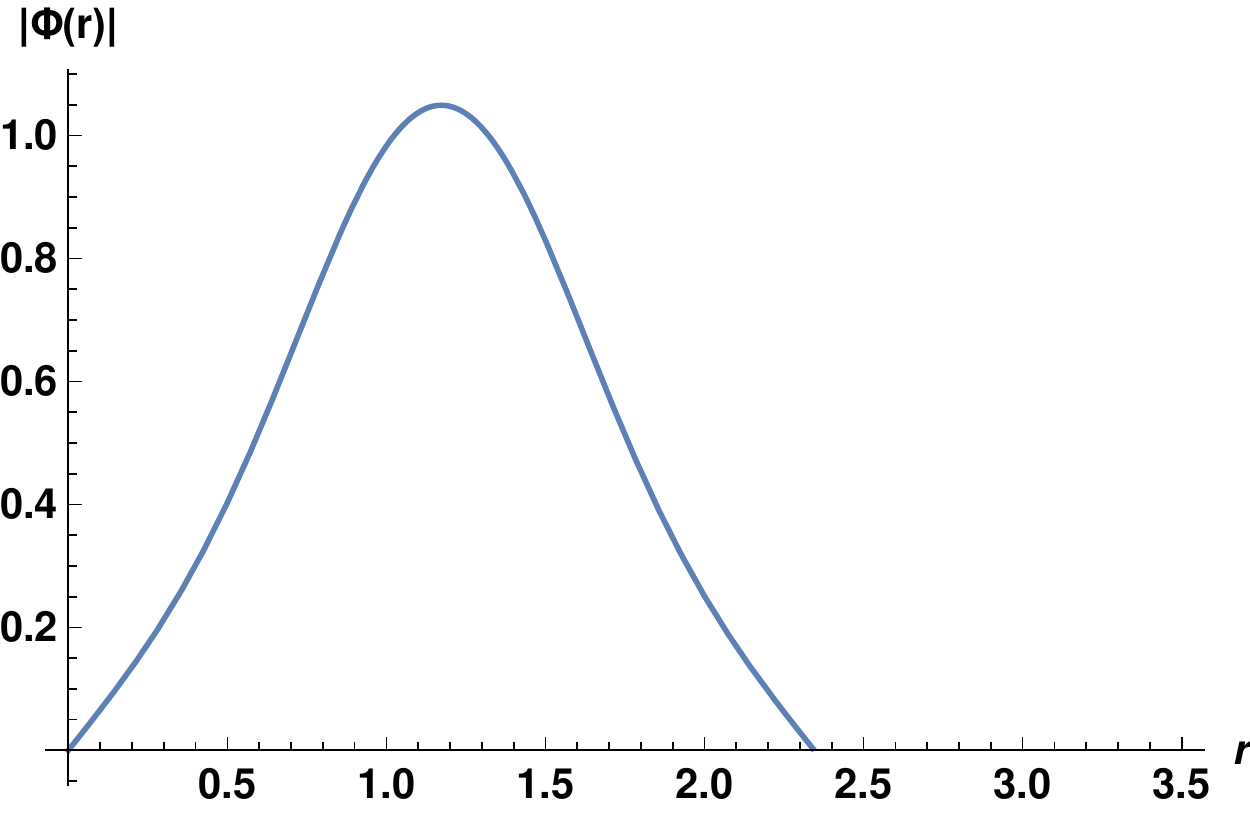}}
}\\
\caption{\small Depicted on the left column are all possible potentials (up to sextic order) admitting compact vortex solutions; on the right a characteristic solution (\emph{vortex compacton}) for each potential. For each solution the field is exactly zero for $r\geq r_1$, where $|\Phi(r_1)|=0$ with $r_1\neq 0$.
\label{fig:compactonsolutions}}
\end{figure}

We conclude this section by showing a numerical computation of all possible compacton solutions in Figure~\ref{fig:compactonsolutions}.

\Comment{
\subsection{Summary of all possible vortex solutions with renormalizable potentials}

We now summarize all the vortex solutions we found in terms of the shape of the potential they arise from. For sextic potentials 
\be
V= m^2 |\Phi|^2 + \lambda |\Phi|^4 + C_1 |\Phi|^6
\ee
there are basically three distinct shapes depending on all possible values for the parameters $m^2$, $\lambda$, and $C_1$. Figure~\ref{fig:shapes} schematically shows the possible shapes for bounded-from-below potentials. The solutions that arise from potentials that are unbounded from below simply correspond to the inverted versions of these shapes.  

\begin{figure}[ht!]
\begin{center}
\includegraphics[scale=0.26]{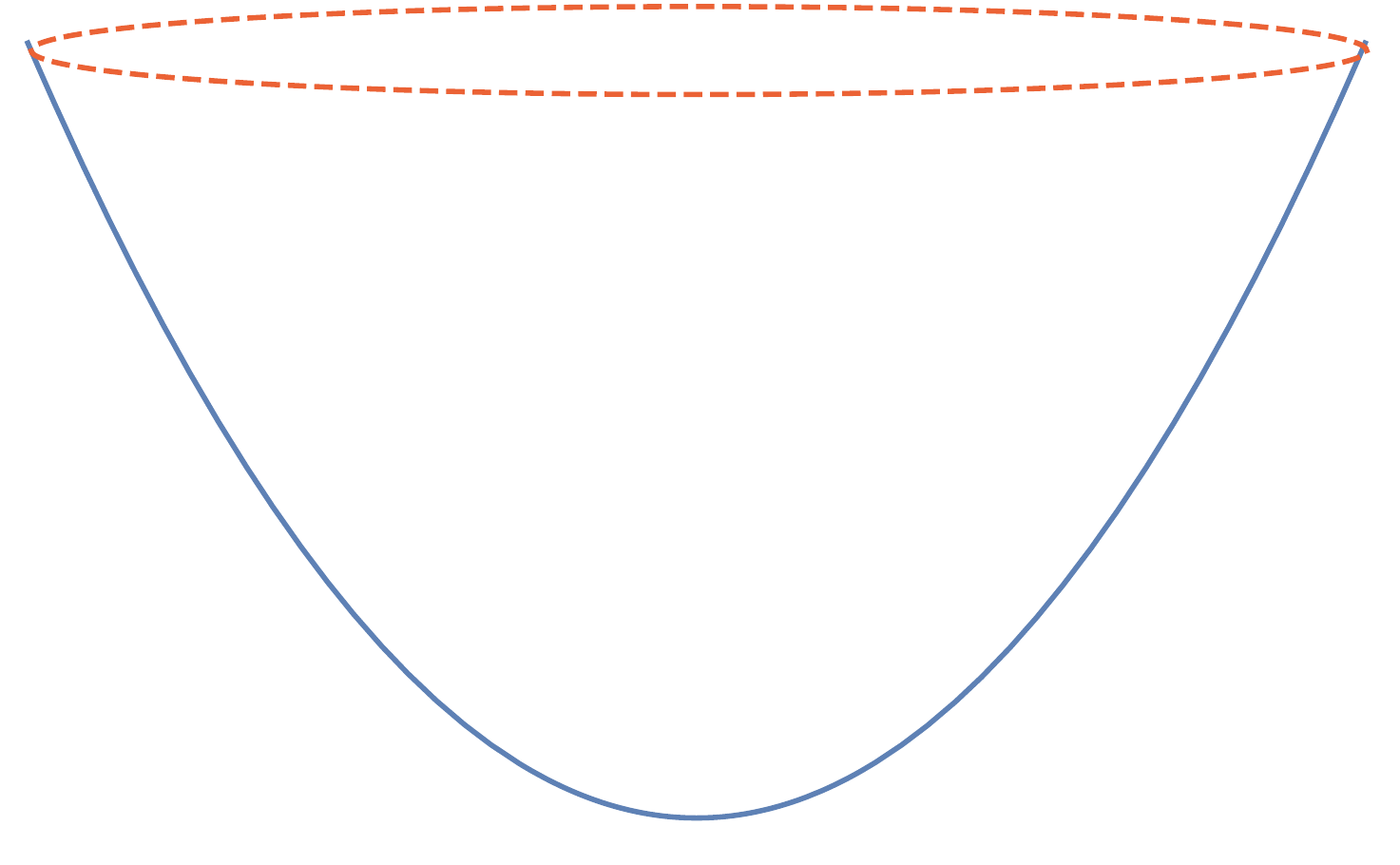} \hspace{10pt} \includegraphics[scale=0.30]{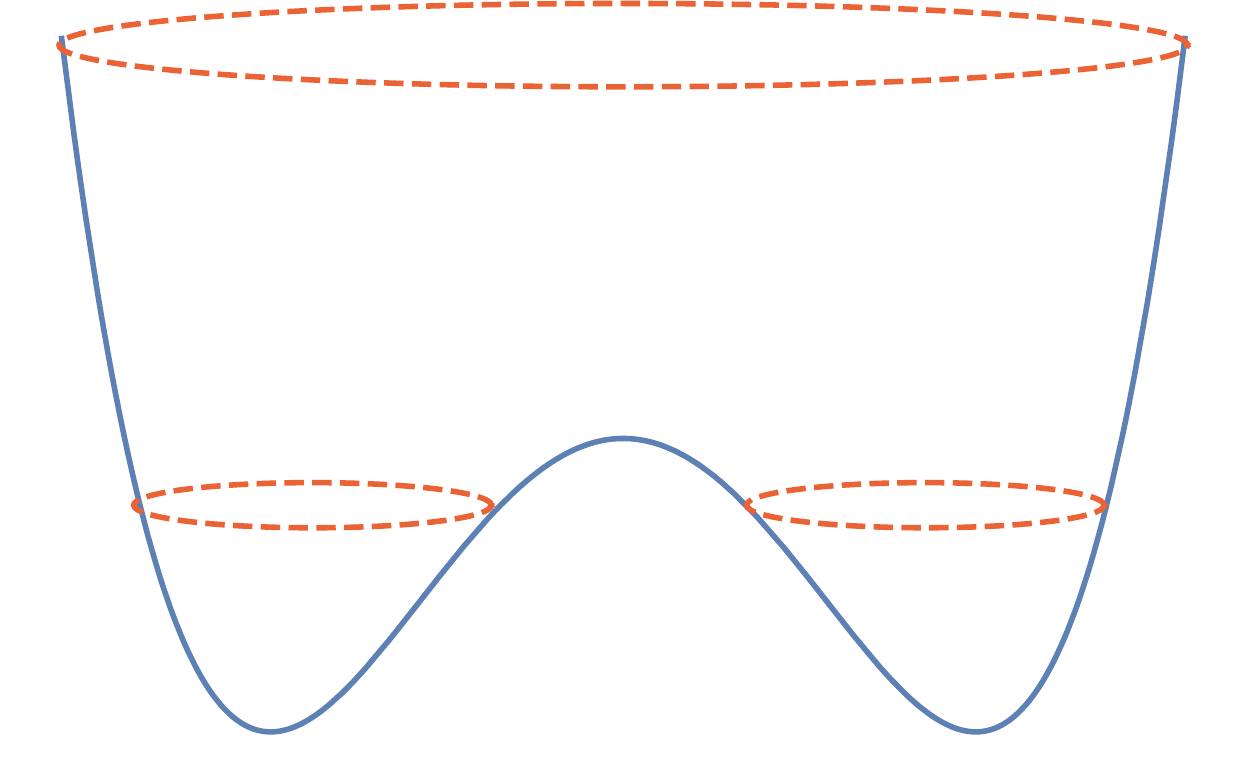} \hspace{10pt} \includegraphics[scale=0.30]{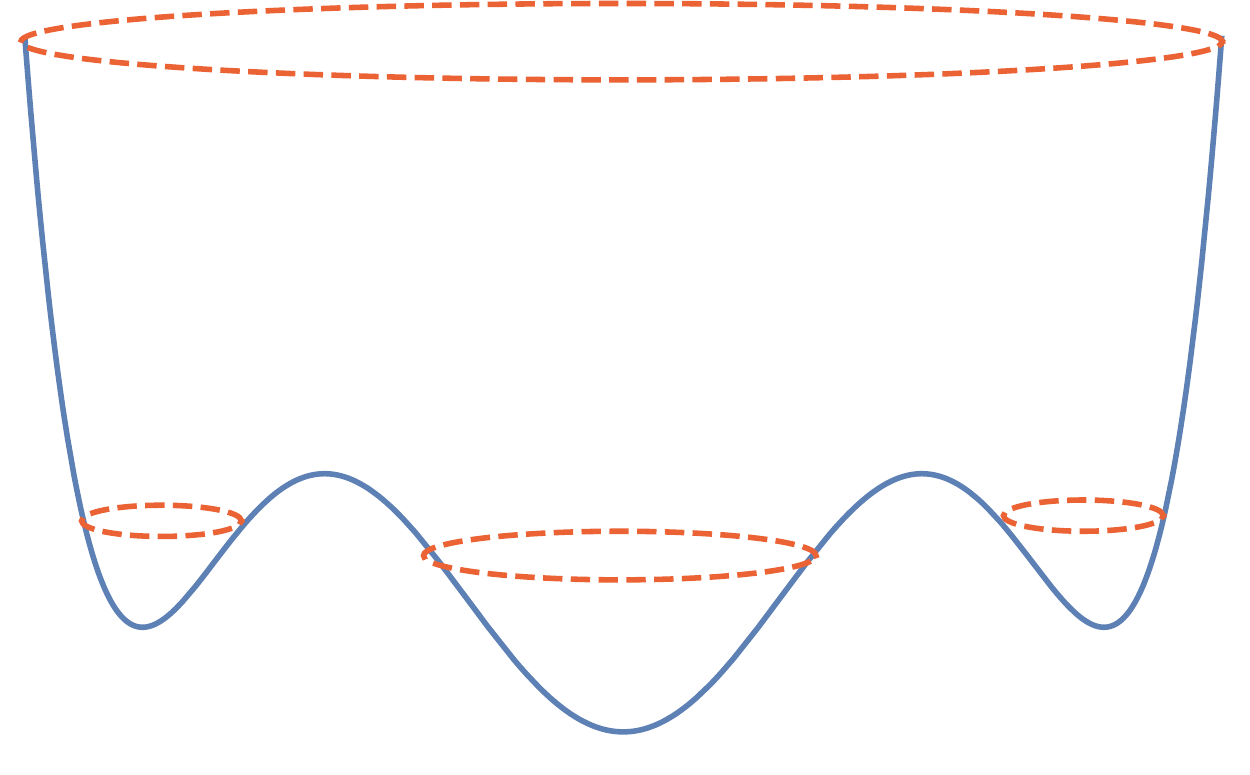}
\caption{Schematic shapes for the potential $V= m^2 |\Phi|^2 + \lambda |\Phi|^4 + C_1 |\Phi|^6$ with $C_1>0$.}
\label{fig:shapes}
\end{center}
\end{figure}
\noindent We will loosely refer to each shape as single, double, and triple vacuum shapes. The following list exhausts all possible cases:
\begin{enumerate}
\item{\emph{Single-vacuum:}} 
	\begin{itemize}
		\item $\lambda \geq 0$ and $m^2 \geq 0$, or
		\item $\lambda^2 < 3 m^2 C_1 $ (with $\lambda <0$ and $m^2>0$)
	\end{itemize}

\item{\emph{Double-vacuum:}} 
	\begin{itemize}
		\item $m^2 < 0$, $\forall \lambda$ or
		\item $m^2=0$ and $\lambda <0$
	\end{itemize}

\item{\emph{Triple-vacuum:}} 
	\begin{itemize}
		\item $\lambda^2 > 3 m^2 C_1 $ (with $\lambda <0$ and $m^2>0$)
	\end{itemize}

\item{\emph{The special case $\lambda = - \sqrt{3 m^2 C_1}$}}: This is the potential arising in the ``massless Higgs'' cases. It has only a trivial vacuum at $|\phi|=0$ (see Figure~\ref{fig:criticalV}). At the location $|\Phi_0|$ where the first derivative of $V$ vanishes, the second derivative of $V$ also vanishes, thus yielding a zero mass for the putative Higgs. 

\end{enumerate}

\begin{figure}
\begin{center}
\includegraphics[scale=0.4]{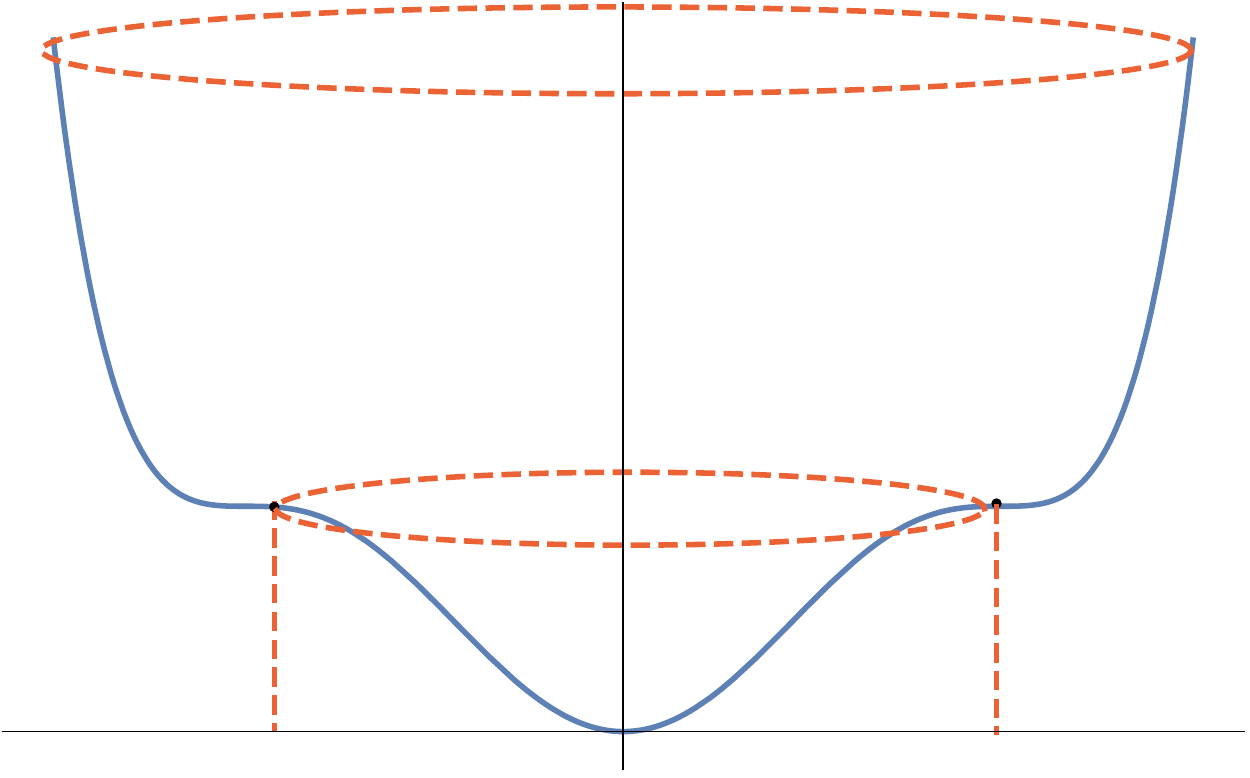}
\caption{The lower dashed lines indicate the location where the slope vanishes where the potential is neither a local minimum nor a local maximum.}
\label{fig:criticalV}
\end{center}
\end{figure}
}

\subsection{Nonrenormalizable potentials and vortices in the FQHE}

{\bf $m$ vortex}

Having in mind the application to FQHE, where as we saw the Laughlin wavefunction (\ref{laughlin}) for $N=1$ (one electron)  and filling fraction $
\nu =1/m$ is of the type
\be
|\Phi(r)|=Ar^m e^{-\a r^2}\;,
\ee
with $\a>0$, we want to find nonrenormalizable potentials that would have it as a solution. 

We first notice however that, while we have the characteristic $m$-vortex behaviour $|\Phi(r)|\sim A r^m$ at $r\rightarrow 0$, the fact that the solution starts and 
ends at $\Phi=0$ means that it is not a vortex of the type we had considered. It is also not a compacton solution, which has a compact domain, and is 
symmetric with respect to some $r_0$. 

Rather, the simplest explanation that still allows us to think of the wavefunction as a soliton solution, is to 
consider that the potential changes between the low $r$ and high $r$ domains. That is actually sensible if we think of the potential $V(\Phi)$ as 
some effective potential in a quantum theory, that would a priori depend on the scale. Then at $r\rightarrow 0$ we would get the UV potential, which 
for asymptotically free theories would be the fundamental one, while are $r\rightarrow \infty$ we would get the IR potential, which would be the 
renormalized one. Since in both cases, we only probe small $|\Phi|$'s, only the $|\Phi|\rightarrow 0$ limit of the potential is relevant.

It is also not entirely clear what would be 
the interpretation of the scalar field in relation to the wavefunction of electrons in FQHE. But later, we will take a nonrelativistic limit to 
obtain a cousin of the Jackiw-Pi model, which can be thought of as describing a nonlinear Schr\"{o}dinger equation. We will therefore
extend to this relativistic case the intepretation of $\Phi$ as a wavefunction, in order to consider the Laughlin wavefunction as a 
vortex solution.

{\bf The IR potential and the pure gaussian}

In order to understand the IR potential, we consider the wavefunction at $r\rightarrow\infty$, where the Gaussian factor dominates. Therefore we 
study the potential arising from a pure gaussian, which is a monotonically decreasing function, so it could correspond to a normal vortex solution of 
the type we studied in this paper. 

We consider then the wave function
\be
|\Phi|=e^{-\a r^2}\Rightarrow \ln |\Phi|=-\a r^2\;,
\ee
Then, by substituting in the equations of motion, we obtain
\be
\frac{|\Phi|''}{|\Phi|}=4\a^2 r^2-2\a=-2\a(\ln|\Phi|^2+1)=\frac{dV}{d|\Phi|^2}\;,
\ee
which gives the {\em exact } solution for the potential
\be
V(|\Phi|)=-2\a|\Phi|^2\ln |\Phi|^2.
\ee

Note that this potential,
\be
V(x)=-2\a x\ln x\;,
\ee
satisfies $V(0)=0$ (as a limit), $V(1)=0$, $V(1/e)=2\a/e$, and its derivative
\be
V'(x)=-2\a(1+\ln x)
\ee
satisfies $V'(0)=+\infty$, $V'(1/e)=0$. Therefore the potential starts at zero, goes to a maximum at $x=1/e$, then goes back to zero at $x=1$, after
which it stays negative. 

However, as we said, the only relevant part of the potential is at small enough $|\Phi|$. 

{\bf The UV potential and the power law}

On the other hand, in the UV, or $r\rightarrow 0$ limit, the wave function is approximated by the power law
\be
|\Phi|= r^m\;,
\ee
which from the equation of motion 
\be
|\Phi|''=\frac{dV}{d|\Phi|}\;,
\ee
means that the potential is 
\be
V(|\Phi|)\simeq \frac{m^2}{2}[|\Phi|^2]^{1-\frac{1}{m}}\;,
\ee
however, as before, it is valid only in the $|\Phi|\rightarrow 0$ limit. 

{\bf Interpretation}

We see that the IR potential is universal, and is 
\be
V(|\Phi|^2)=-2\a|\Phi|^2\ln|\Phi|^2(+{\cal O}(|\Phi|^4)).\label{leadlog}
\ee
In fact, if we would have a general renormalizable potential like in the previous subsection, but we would consider a one-loop Coleman-Weinberg 
type calculation of the quantum effective potential, with a general renormalization condition, we expect that the quadratic (mass) term gets 
logarithmic corrections.\footnote{In 4 dimensions, the standard Coleman-Weinberg 1-loop calculation at nonzero renormalized mass $m_{\rm ren}$ 
gives a correction to $m^2_{\rm ren}$ of 
\be
m^2_{\rm ren}\frac{\lambda_{\rm ren}}{(8\pi)^2}\ln\left(1+\frac{\lambda_{\rm ren}\bar\phi^2}{2m^2_{\rm ren}}\right).\nonumber
\ee
(see eq. 16.2.15 in \cite{Weinberg:1996kr}).} 
However, these log corrections actually vanish at $\bar\phi=0$, instead of becoming infinite. But if one considers instead a Wilsonian effective 
action like the one considered in the case of ${\cal N}=2$ supersymmetric gauge theory by Seiberg \cite{Seiberg:1988ur}, one does obtain 
such corrections. In the ${\cal N}=2$ case, Seiberg obtained a correction to the effective action that amounts to a replacement of the Wilsonian 
gauge coupling factor $\frac{1}{e^2(\mu)}$, with $\mu$ the Wilsonian scale, by 
\be
\frac{1}{e^2(\langle \phi\rangle)}=\frac{1}{e^2(\mu)}+\frac{3}{4\pi^2}+\frac{1}{4\pi^2}\ln\frac{\langle \phi\rangle^2}{\Lambda^2}.
\ee
Likely a Wilsonian approach for a 3 dimensional renormalizable scalar theory will lead to a correction of the type in (\ref{leadlog}).
Then at small $|\Phi|$, indeed the leading term in the potential is (\ref{leadlog}).

On the other hand, the UV potential is non-universal, $V\sim |\Phi|^{2-2/m}$, but becomes universal at $m\rightarrow \infty$,
\be
V(|\Phi|^2)\sim |\Phi|^2(+{\cal O}(|\Phi|^4))\;,
\ee
which would be indeed the leading renormalizable term (mass term) at $|\Phi|\rightarrow 0$. It is unclear why we would need to consider the $m\rightarrow
\infty$ to make sense of this situation, and at finite $m$ the potential is non-analytic in $|\Phi|^2$. 

\section{Non-Abelian vortex solutions}

We now look for nonabelian vortex solutions. We will consider the scalar field $\phi$ in the adjoint representation of SU(N). With hermitian 
generators $t^a$ obeying $[t^a,t^b]= i f^{abc}t^c$ with structure constants $f^{abc}$, we use the normalization $Tr(t^at^b)=\delta^{ab}$. In 
the adjoint representation we can write the matrix field $\phi$ as $\phi = \phi_a t^a$, thus, we have $N^2-1$ complex scalars $\phi_a$.  We 
now look for vortex solutions starting from the action
\be 
\label{nonabelianL}
S =\int d^3x \left( -\tr(D_\mu \phi^\dagger D^\mu \phi) - \frac{k}{2\pi} \epsilon^{\mu\nu\rho} a_\mu \partial_\nu A_\rho + \frac{1}{e}A_\mu 
j^\mu_{\rm vortex}(t) - V\right),
\ee 
with covariant derivative $D_\mu \phi \equiv (\partial_\mu -i e a_\mu)\phi$ and potential
\eal{
\label{nonabelianpot}
V =& m^2 \tr(\phi^\dagger \phi) + \lambda_1 \tr^2(\phi^\dagger \phi)+ C_1\tr^3(\phi^\dagger \phi) +\lambda_2 \tr([\phi^\dagger,\phi]^2) \\
& + C_2  \tr (\phi^\dagger \phi) \tr ([\phi^\dagger,\phi]^2)
+C_3 \tr [\phi^\dagger,[\phi,\phi^\dagger]][[\phi,\phi^\dagger],\phi].
}
Note that we perform a non-abelian extension only on the scalar fields; the gauge fields $a_\mu$ and $A_\mu$ are still abelian. The main idea 
we pursue in this section is to \emph{recycle} the abelian solutions that we have already found in section 3 by constructing nonabelian ones out of  them. 

\subsection{Equations of motion}
The variation of $A^\mu$ yields the same equation as in the abelian case, \emph{i.e.,}
\eal{
\epsilon^{\mu\nu\rho} \partial_\nu A_\rho = \frac{2\pi}{ke}j^\mu_{\rm vortex}(t).
}
Varying $a^\mu$ gives  
\eal{
\tr \left(\phi(D_\mu \phi)^\dagger-\phi^\dagger D_\mu \phi\right) = \frac{k}{2\pi ie } \epsilon^{\mu\nu\rho} \partial_\nu A_\rho,
}
thus, as explained in section 3.1, setting $A_\mu =0$ \emph{after} writing the equations of motion leads to the generalization
\eal{
\tr \left(\phi(D_\mu \phi)^\dagger-\phi^\dagger D_\mu \phi\right) =0
}
Using $\phi = \phi_a t^a$, this condition imposes
\eal{
\label{eom2}
\phi_a (D_\mu \phi_a)^* - \phi_a^* (D_\mu \phi_a)=0 \qquad \text{(no sum over $a$)}
}
for each field component $\phi_a$, where $D_\mu \phi_a \equiv \partial_\mu \phi_a - i e a_\mu \phi_a$. As is usual when looking for vortex solutions,  it is helpful to separate the magnitude and phase of the complex fields as $ \phi_a = |\phi_a|e^{i \theta_a}$. Assuming the radial ansatz $|\phi_a(r)|$ and $\theta_a(\varphi)$ where $r$ and $\varphi$ are respectively the radial and polar-angle coordinates, condition \eqref{eom2} yields
\be
\label{puregaugecond}
\partial_\mu \theta_a = e a_\mu. 
\ee
Note that the right-hand side does not depend on the index $a$ (the Chern Simons field $a_\mu$ is just an abelian gauge field), 
thus all angles $\theta_a$ only differ from one another by a simple constant of integration $c_a$. Since there are as many equations as $\phi_a$ fields are, we also have as many integration constants,  thus
\eal{
\theta_a(\varphi) = \theta(\varphi) + c_a
}
where 
\be
 \theta(\varphi)  \equiv e \int_0^\varphi a_{\varphi}(\varphi') d\varphi' 
\ee
The main new ingredient comes from the nonabelian potential above. Under $\delta \phi^\dagger$ we obtain the field equation
\eal{
D_\mu D^\mu \phi =& m^2 \phi + 2 \lambda_1 \phi  \tr (\phi^\dagger \phi) + 3 C_1 \phi \tr^2(\phi^\dagger \phi) + 2 \lambda_2 [\phi,[\phi^\dagger, \phi]] + C_2 \phi \tr\left([\phi^\dagger,\phi]^2\right)\\
&+2C_2 \tr(\phi^\dagger \phi)  [\phi,[\phi^\dagger, \phi]] + C_3 \left([[\phi,\phi^\dagger],B] +[[B,\phi^\dagger],\phi]+[\phi,[B^\dagger,\phi]]\right),
}
where $B \equiv [[\phi,\phi^\dagger],\phi]$.

Later on it will be useful to express the commutators in terms of the scalar fields $\phi_a$. For instance, the commutator $[\phi^\dagger,\phi]$ is
\eal{
{}[\phi^\dagger,\phi]_{ab} &= \phi_i^* \phi_j [t^i,t^j]_{ab}\\
&=i |\phi_i| |\phi_j| e^{i(\theta_j-\theta_i)} f^{ijk} (t^k)_{ab},
}
and because $f^{ijk}$ is completely antisymmetric, only the antisymmetric part of $|\phi_i| |\phi_j| e^{i(\theta_j-\theta_i)}$ survives, therefore
\eal{
\label{commutator}
{}[\phi^\dagger,\phi]_{ab} &= \tfrac{1}{2} i |\phi_i| |\phi_j| \left( e^{i(\theta_j - \theta_i)} - e^{i(\theta_j - \theta_i)}\right) f^{ijk} (t^k)_{ab}\\
&=  |\phi_i| |\phi_j| \sin (\theta_i - \theta_j)f^{ijk} (t^k)_{ab}
} 
Thus, if we choose all the phases $\theta_i$ to differ from one another by an integer factor of $\pi$, then all the  components of the matrix $[\phi^\dagger,\phi]$ vanish identically. 

\subsection{Vortex solutions}
%

In polar coordinates we have $D_\mu D^\mu \phi = -D_t^2 \phi + D_r^2 \phi + r^{-2} D_{\varphi}^2 \phi$. For static vortices $\partial_t \phi  =0$ and also, from \eqref{puregaugecond}, we obtain $a_t=a_r=0$. Therefore, $D_t \phi=0$ yielding
\eal{
D_\mu D^\mu \phi =& D_r(\partial_r \phi) + r^{-2}D_{\varphi}(\partial_\varphi \phi - ie a_\varphi \phi) \\
=& D_r \sum_a t^a e^{i \theta_a} \partial_r |\phi_a| + r^{-2}D_\varphi\sum_a i|\phi_a|e^{i\theta_a}(\partial_\varphi\theta_a -e a_\varphi)\\
= & \partial_r \sum_a t^a e^{i \theta_a} \partial_r |\phi_a| \\
= & \sum_a t^a e^{i \theta_a} |\phi_a|'' 
}
where in going from the second line to the third we again used \eqref{puregaugecond}. Here $' \equiv \partial/\partial r$.

For simplicity we will consider the case where $[\phi^\dagger,\phi]=0$. From \eqref{commutator} we have
\be
{}[\phi^\dagger,\phi] =|\phi_i| |\phi_j| \sin (\theta_i - \theta_j)f^{ijk} t^k,
\ee
thus, as mentioned above, due to the $\sin (\theta_i - \theta_j)$ factor we can make the commutator vanish by choosing all phases to differ one another by an integer factor of $\pi$ (up to an overall phase). Doing this the equations of motion reduce to
\eal{
\frac{|\phi_1|''}{|\phi_1|} &= m^2 + 2 \lambda_1 (|\phi_1|^2+|\phi_2|^2 +\cdots) + 3C_1(|\phi_1|^2+|\phi_2|^2 +\cdots)^2\\
\frac{|\phi_2|''}{|\phi_2|} &= m^2 + 2 \lambda_1 (|\phi_1|^2+|\phi_2|^2 +\cdots)+ 3C_1(|\phi_1|^2+|\phi_2|^2 +\cdots)^2\\
\vdots  & \hspace{100pt} \vdots \hspace{100pt} \vdots 
}
Considering the ansatz $|\phi_1|=|\phi_2|=\cdots=f(r)$ leads to 
\be
\frac{f''(r)}{f(r)} = m^2 + 2 \lambda_1 (N^2-1)f^2(r) + 3C_1 (N^2-1)^2 f^4(r).
\ee
which is precisely the same equation of motion for the abelian case that we extensively studied in sections 3.1 and 3.2. Thus, the main idea here is that by considering the ansatz $[\phi^\dagger,\phi]=0$ we can simple borrow all the vortex solutions we have already found to construct new non-abelian ones.

For simplicity, let's make only one of the components of $\phi$ to wind, say, $\theta_1$, therefore\footnote{For $SU(2)$ fields in the fundamental representation this solution is known as the (1,0) string. Winding the other component instead is called (0,1) string.}
\eal{
\phi_1 = f(r) e^{i\theta}\,, \quad \phi_2=\phi_3= \cdots = f(r)
}
We can also find the topological charge defined as
\be
 T =  \tr \oint \Omega_i dx^i 
\ee
where the matrix $\Omega_i$ is
\be
\Omega_i[\phi] \equiv \frac{1}{i} (\partial_i \phi) \phi^{-1}
\ee
For the case of SU(2) we have
\be
\phi = \frac{1}{\sqrt{2}} \begin{pmatrix}
    \phi_3 & \phi_1-i\phi_2\\
    \phi_1+i \phi_2 & -\phi_3
  \end{pmatrix}
\ee
which, for our solution, becomes
\be
\phi = \frac{f(r)}{\sqrt{2}} \begin{pmatrix}
    1 & e^{i\theta}-i\\
     e^{i\theta}+i& -1
  \end{pmatrix}
\ee
For a $n$-vortex solution with all vortices on top of each other at the origin, we have $\theta=n \varphi$ and 
\eal{
\Omega_\varphi = \frac{1}{i}(\partial_\varphi \phi) \phi^{-1} =  \frac{ne^{in\varphi}}{2+e^{2in\varphi}} 
\begin{pmatrix}
e^{in\varphi} + i & -1\\
1 & e^{in\varphi} - i
\end{pmatrix}
}
thus $\tr \Omega_\varphi = 2n e^{2in\varphi}(2+e^{2in\varphi})^{-1}$. The topological charge is then
\eal{
T &= \tr \int_0^{2\pi} \Omega_{\varphi} d\varphi =  2n \int_0^{2\pi} \frac{e^{2in\varphi}}{(2+e^{2in\varphi})}d\varphi = 0
}
\emph{i.e.} it vanishes. Note that for a trivially Abelian solution $\phi =e^{in\varphi} \one$, we would get a nonzero charge, as we would for 
a more general non-Abelian solution, so this seems to suggest that the solution is somehow nontrivial.

\section{Non-relativistic limit}

We want to take a nonrelativistic limit on the Abelian model for the vortex with source, since we know that the nonrelativistic limit of the 
Landau-Ginzburg model gives the very interesting Jackiw-Pi model with vortices, so we expect something similarly interesting 
to happen in our case. We consider therefore as a starting point the relativistic Lagrangean
\be
\mathcal{L} = \frac{\kappa}{2\pi} \epsilon^{\mu\nu\rho} a_\mu \partial_\nu A_\rho -| D_\mu \phi|^2 + \frac{1}{e} A_\mu j^\mu_{\rm vortex}- V(|\phi|)\;,
\ee
where $D_\mu=\d_\mu -i a_\mu$ and the potential is
\be
V(|\phi|) = m^2 |\phi|^2 + \lambda |\phi|^4 + C_1 |\phi|^6. 
\ee
Note that we could consider a priori higher powers of $|\phi|$ in $V$, giving nonrenormalizable potentials. But as we will see shortly, already the 
sextic term vanishes in the nonrelativistic limit, and the same will be true for higher powers. 

In order to take the nonrelativistic limit, we reintroduce factors of $\hbar $ and $c$ by dimensional analysis, imposing that $[\lambda]=M$ and 
$[C_1]=0$ (dimensionless). We also write $\d_\mu=(\d_t/c,\d_i)$ and $a_\mu=(a_t/c,a_i)$, $A_\mu=(A_t/c,A_i)$.
We obtain\footnote{Since $[m]=M, c=L T^{-1}$ and $[S]=[\hbar ]=M L^2 T^{-1}$, we get $[{\cal L}]=M T^{-2}$. 
Then the kinetic term $-(\d_i\phi)^2/2$ means that $[\phi]= M^{1/2} L T^{-1}$ and $D_\mu =\d_\mu -i a_\mu$ means that $[a_\mu]=L^{-1}$, and 
similarly we take $[A_\mu]=L^{-1}$.}
\be
\mathcal{L} = \frac{\kappa(\hbar c)}{2\pi} \epsilon^{\mu\nu\rho} a_\mu \partial_\nu A_\rho +\frac{1}{c^2}| D_t \phi|^2 -|D_i\phi|^2
+ \frac{(\hbar c)}{e} A_\mu j^\mu_{\rm vortex}-\frac{m^2c^2}{\hbar^2}|\phi|^2 -\frac{1}{\hbar^2} \lambda |\phi|^4 - \frac{C_1}{\hbar^2c^2} |\phi|^6
\ee
We have multiplied $\kappa$ by $(\hbar c)$ in order that $\kappa$ is kept dimensionless (a number), but we can reabsorb $k=\kappa (\hbar c)$. 
Similarly, we have considered $j^0_{\rm vortex}=e\delta^2(x)$, but we could absorb $(\hbar c)$ in $j^\mu_{\rm vortex}$. 

In order to perform the nonrelativistic limit, we redefine the field as (see for instance \cite{Lopez-Arcos:2015cqa})
\be
\label{eq:fieldredef}
\phi(t,\vec x)= \frac{\hbar}{\sqrt{2m}} e^{-i\frac{mc^2}{\hbar}t}\psi(t,\vec x)
\ee
Note that we could add to this formula a complex conjugate part (with a different field $\hat\psi^*$) for the antiparticle sector, 
but in the nonrelativistic limit the particle and 
antiparticle sectors are separately conserved, so we can keep only the particles. 

When replacing the field in the action and taking the $c\rightarrow \infty$ limit, the mass term in the potential $\propto (mc/\hbar)^2$ will cancel 
due to the redefinition of the field \eqref{eq:fieldredef} (via $\d_t$ acting on $e^{-i\frac{mc^2}{\hbar}t}$), the sextic term in the potential will vanish, being $\propto 1/c^2$, 
and we are left only with the quartic term in the potential. Also the terms $\hbar^2/2mc^2[\d_t\psi^*\d_t \psi
- i a_0 \psi \partial_t \psi^* + i a_0 \psi^* \partial_t \psi + a_0^2 |\psi|^2]$, coming from $|D_t\phi|^2/c^2$, vanish in the limit, leading finally to the 
Lagrangean 
\be
\mathcal{L}_{NR} = \frac{\kappa(\hbar c)}{2\pi} \epsilon^{\mu\nu\rho} a_\mu \partial_\nu A_\rho + \frac{(\hbar c)}{e} A_\mu j^\mu_{\rm vortex} 
+ i\hbar  \psi^*D_t\psi - \frac{\hbar^2}{2m}  | \vec D \psi|^2 - \frac{\lambda\hbar^2}{(2m)^2 } |\psi|^4\label{nrmodel}
\ee
For comparison, the Jackiw-Pi Lagrangean at $\hbar=1$ is \cite{Jackiw:1990tz}
\be
\mathcal{L}_{\rm Jackiw\!\!-\!\!Pi} = \frac{k}{2} \epsilon^{\mu\nu\rho} A_\mu \partial_\nu A_\rho 
+ i \psi^*D_t\psi - \frac{1}{2m}  | \vec D \psi|^2 + \frac{g}{2} |\psi|^4 \;,
\ee
where $k=\kappa c$, and $D_\mu=\d_\mu - iA_\mu $. So the difference from the Jackiw-Pi model is the source term, as well as the 
gauge fields (we have a mixed CS term instead of the electromagnetic one, and an $a_\mu$ in the covariant derivative instead of $A_\mu$). 
We would also need $\lambda<0$, i.e. a negative definite quartic potential to have the same Lagrangean.

Note that the Jackiw-Pi model can be thought of as describing a (gauged) nonlinear Schr\"{o}dinger equation (the equation for $\psi$ takes this form, 
see also below), hence we can think of $\psi$ as the wavefunction for some system. We have in fact already used this interpretation in the 
relativistic case, in order to map to the Laughlin wavefunction. In the nonrelativistic model (\ref{nrmodel}), the sextic term in the potential vanished 
since it came with $1/c^2$. It is easy to see that any new power of $|\phi|$ in the potential
comes with an extra $1/c$ factor (since $[\phi]=M^{1/2}L T^{-1}$), so all higher terms will vanish in the nonrelativistic limit, and the model (\ref{nrmodel})
is the result even for a general nonrenormalizable relativistic potential!

The equations of motion for the new nonrelativistic model (\ref{nrmodel}) we have obtained are
\bea
&&i D_t \psi = -\frac{2}{2m} D_i^2 \psi + \frac{\lambda}{2m^2} |\psi|^2 \psi \\
&&\frac{\kappa e}{2\pi}\epsilon^{\mu\nu\rho}\partial_\nu a_\mu = j^\rho_{\rm vortex}(t)\\
&&i\left[\psi^* D_i \psi - \psi(D_i \psi)^*\right] =  \frac{\kappa m}{\pi} \epsilon^{i\nu\rho}\partial_\nu A_\rho\\
&&\frac{\kappa}{2\pi} B = - |\psi|^2\label{nreqs}
\eea

\subsection{Nonrelativistic vortex}

We want to consider a solution to (\ref{nreqs}) of the type vortex with source considered in the rest of the paper. 

As before, we look for vortex solutions with $\psi=|\psi| e^{i\theta}$ and 
\be
a_\mu=\d_\mu\theta\;,
\ee
where $\theta$ is the polar angle, specifically $a_0=\d_0\theta=0$ and $a_i=\d_i\theta$. This solves the second equation exactly, as before, if $
\kappa=2$. It also solves the third equation, if we put $\epsilon^{i\nu\rho}\d_\nu A_\rho=0$ through $A_0=0,\d_0 A_j=0$. 
Then the fourth equation simply defines the magnetic field from $\psi$,
\be
\frac{\kappa}{2\pi}B=\frac{\kappa}{2\pi}\epsilon^{ij}\d_i A_j=-|\psi|^2.
\ee
The only equation left to solve is the first equation, which in the static case ($\d_t \psi=0$) becomes
\be
|\psi|''=\frac{\lambda}{2m}|\psi|^3.\label{psidprime}
\ee

This is solved in a manner exactly analogous to the case of the purely sextic potential reviewed in section 3.2 and done in \cite{Murugan:2014sfa}.
We define $v=|\psi|'$, and then we can rewrite the equation of motion as 
\be
v\frac{dv}{d\psi}=\b |\psi|^3\;,
\ee
where $\b\equiv \lambda/(2m)$. This is easily integrated to give 
\be
\frac{v^2}{2}=\frac{\b |\psi|^4}{4}+K_1\Rightarrow |\psi|'=\pm \sqrt{\frac{\b }{2}|\psi|^4+2K_1}\;,\label{psiprime}
\ee
which means that the formal solution for $\psi(r)$ is given implicitly by 
\be
r+K_2=\pm \int \frac{d|\psi|}{\sqrt{\frac{\b}{2}|\psi|^4+2K_1}}.
\ee

If $|\psi|$ is small and $K_1\neq 0$, we can write the solution as
\be
r+K_2\simeq \pm\frac{|\psi|}{\sqrt{2K_1}}\rightarrow |\psi|(r)\simeq \pm \sqrt{2K_1}(r+K_2)\;,
\ee
whereas if $|\psi|$ is large or $K_1=0$, we have 
\be
r+K_2\simeq \mp \sqrt{\frac{2}{\b}}\frac{1}{|\psi|}\Rightarrow |\psi|(r)\simeq \mp \sqrt{\frac{2}{\b}}\frac{1}{r+K_2}.
\ee

In order to have a solution that goes to zero at $r=0$, we see that we need $K_1>0$, $K_2=0$ and the plus sign in the formal solution, 
at least close to zero. But then in the general solution, from (\ref{psiprime}), $|\psi|$ would increase without bound 
(thus having a solution with infinite total energy) unless $\b<0$, i.e. $\lambda<0$.
This kind of solution was discarded in the case of the purely sextic potential in  \cite{Murugan:2014sfa} because it corresponds to a potential that 
is unbounded from below. But in the current case, we have a non-relativistic theory that could still have a relativistic (UV) completion with 
$C_1>0$, so a full potential bounded from below, yet in the nonrelativistic limit have negative definite potential. Moreover, this is also the case 
of the Jackiw-Pi model, as we saw, so now we need to consider this possibility. 

Before we continue, let us note that the alternative is to consider $\b>0$, but then for a solution that goes to zero at infinity (thus has a finite energy), 
$|\psi|(r=0)$ must be nonzero. This is the analog of the case (\ref{phiatzero}) from the purely sextic case. Now also we can choose 
$K_1=0$, keeping $K_2>0$, and obtain the solution 
\be
|\psi|(r)=+\sqrt{2}{\b}\frac{1}{r+K_2}\;,
\ee
which would have finite energy, since it goes to zero asymptotically, but is ill-defined at $r=0$ (and is not a solution at $r=0$). It is not 
clear if this would make sense for the vortex solutions with source. 

{\bf Compacton solution}

But we can certainly consider further the solution in the case $\b<0$ ($\lambda<0$), $K_1>0$. In this case, the solution starts 
with $|\psi|'>0$ from (\ref{psiprime}), but with $|\psi|''<0$ from (\ref{psidprime}), so $|\psi|'$ goes down, until $|\psi|'=0$, at a maximum
of $|\psi|$. But in order to continue past this point, we need to {\em glue another solution}, with $|\psi|'<0$, i.e. the solution of (\ref{psiprime}), 
but with the negative sign. Also, it must now start at the same value of $r=r_0$, where 
\be
\label{psimax}
|\psi(r_0)|=(-4K_1/\b)^{1/4}\equiv |\psi|_{\rm max}\;,
\ee
and $|\psi|'=0$, and go in the opposite direction. In other words, we must glue the function $|\psi|(2r-r_0)$ to the function $\psi(r)$. 

\begin{figure}[h]
  \centering
   \includegraphics[scale=0.7]{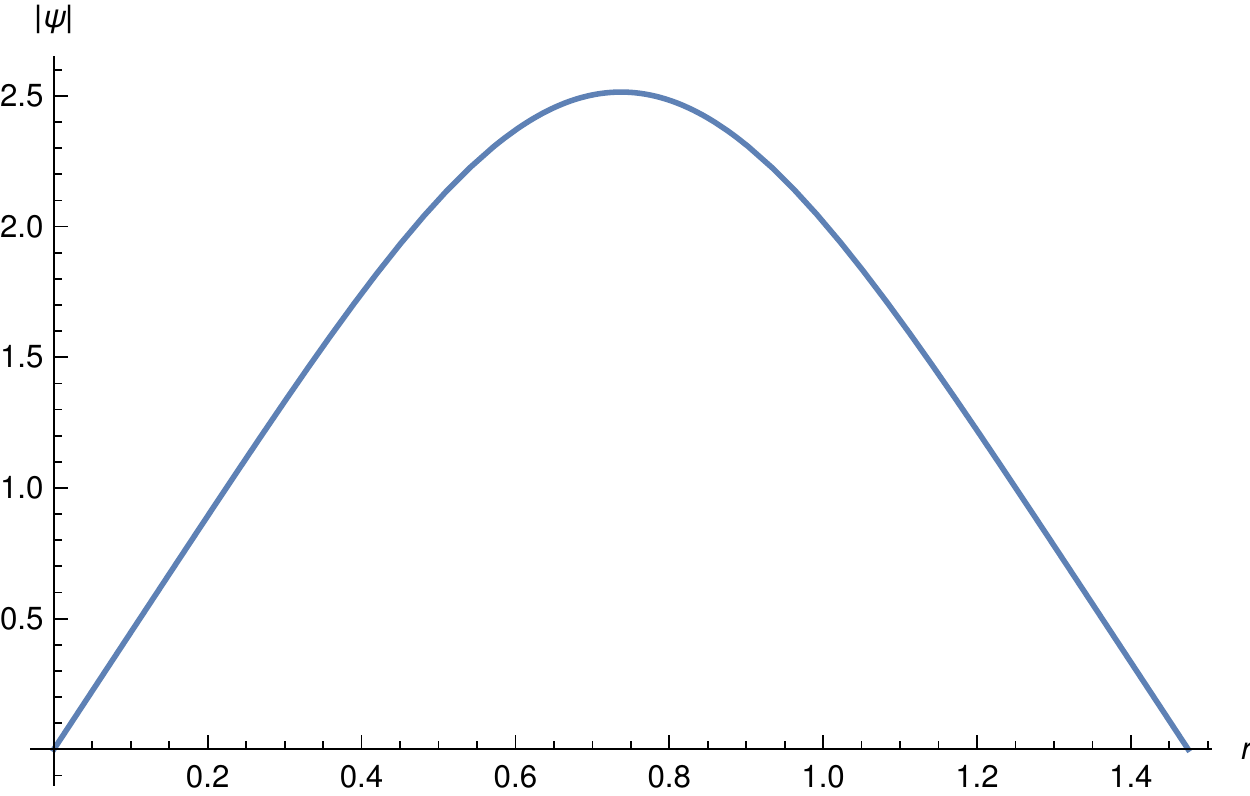}
   \caption{Vortex solution with compact support (\emph{compacton}) for the non-relativistic limit. Here $|\beta|=1$, $K_1=10$. Equations \eqref{psimax} and \eqref{midpoint} yield the numbers $r_0=0.74$ and $|\psi|_{max}=2.51$ which can be seen here.}
   \label{fig:compacton}
\end{figure}

Alternatively, we can simply note that there's a solution in terms of elliptic functions:
\be
\label{eq:compacton}
|\psi(r)|=\sqrt{2}\left(\frac{K_1}{|\b|}\right)^{1/4}{\rm sn}(ar|-1)=|\psi|_{\rm max}{\rm sn}(ar|-1)\;,
\ee
where $a=(K_1|\b|)^{1/4}$ and ${\rm sn}(u|k)$ is the Jacobi Elliptic function with elliptic modulus $k$. One can easily check that \eqref{eq:compacton} satisfies $|\psi|''=\beta |\psi|^3$ for $\beta<0$. The point where it reaches a maximum is 
\be
\label{midpoint}
r_0=\frac{\sqrt{\pi\Gamma(5/4)}}{\Gamma(3/4)}(K_1|\b|)^{-1/4}.
\ee
A plot of this function is in Fig~\ref{fig:compacton}.
We continue the function trivially past $r=2r_0$, by putting $|\psi(r)|=0$ for $r>2r_0$. 

Note that at first sight there seem to be several things wrong with this solution. The derivative at $r=2r_0$ is discontinuous. At $r=2r_0-\epsilon$, since 
$|\psi|=0$, it means from (\ref{psiprime}) that $|\psi|'(2r_0-\epsilon)=-\sqrt{2K_1}\neq 0$, whereas at $r=2r_0+\epsilon$, $|\psi|'(2r_0+\epsilon)=0$. 
That sounds like it should be a problem, but since $|\psi|$ itself is continuous, and so is $|\psi|''$ which vanishes because of the equation of motion 
(\ref{psidprime}), there is no problem, as the equations of motion are satisfied. Note that this means that 
\be
|\psi|(r=2r_0-\epsilon)\simeq -\sqrt{2K_1}(r-2r_0)+{\cal O}((r-2r_0)^3).
\ee
Also, the solution is defined in a finite domain $r=[0,2r_0]$, and is zero outside it, which is certainly unusual, and one could think that it could 
be a problem. 
But there are in fact known solitonic solutions with discontinuous first derivative at one or more points, and these are called ``peakons'' \cite{Camassa:1993zz}. There are also known solitonic solutions with compact support (finite domain), called 
compactons \cite{Rosenau:1993zz}, as we have already
seen in the relativistic case. So our solution is a 
vortex, of the peakon and compacton type, simultaneously (though note that strictly speaking, the peakon has the discontinuity at a peak, whereas 
in our case it is at the boundaries). 

The solution has a nontrivial magnetic field, $B=-|\psi|^2$, therefore a nontrivial magnetic flux,
\bea
\Phi_m&=&\int dx^1 \wedge dx^2 F_{12}=\int B dS=-\int 2\pi r\; dr\; |\psi(r)|^2\cr
&=&-4\pi \left(\frac{K_1}{|\b|}\right)^{1/2} \int r\; dr\; [{\rm sn}(ar|-1)]^2\cr
&=&-4\pi \left(\frac{K_1}{|\b|}\right)^{1/2} a^2\int_0^{2ar_0} x\; dx\; [{\rm sn }(x|-1)]^2=-4\pi cK_1\;, 
\eea
where
\be
c=\int_0^{x_0} x\; dx\; [{\rm sn }(x|-1)]^2;\;\;\; x_0=\frac{\sqrt{\pi\Gamma(5/4)}}{\Gamma(3/4)},
\ee
yielding $c \simeq 0.665$.

But magnetic flux is quantized in terms of the fluxon $\Phi_0=h/e$, so we must impose that 
\be
\Phi_m=N\frac{h}{e}\;,
\ee
where $N\in \mathbb{Z}$, leading to a quantization condition for $K_1$,
\be
K_1=-N\frac{h}{4\pi c e}.
\ee
Here $N$ is an integer number describing the flux of the solution, but unlike the case of the usual vortex, it is not a vortex number. The 
solution we exhibit here is still a one-vortex solution, in the sense of having $\psi=|\psi(r)|e^{i\theta}$. For the more general $M$-vortex solution, we would 
have $\psi=|\psi(r)|e^{iM\theta}$. We have now completely determined the parameters of the vortex solutions.

\section{Embedding in the ABJM model and SYM-CS models in 2+1 dimensions}

We have seen that we have these new kinds of vortex solutions in the presence of a source term. But what does that mean in the context of some 
concrete model, like for instance the ABJM model \cite{Aharony:2008ug}, used as a toy model for various 2+1 dimensional condensed matter systems? 

{\bf Adding source by hand}

One possible answer is that we find our starting action in the absence of 
the source term, and the source term needs to be added by hand, just like what we do when we consider electrons or electric-type solutions. 

In fact, it was shown in \cite{Murugan:2014sfa} that the action found in eq. 2.23 in that paper, 
\bea
S&=&-\int d^3x\ \left[\frac{1}{2}|(\d_\mu - iea_\mu)\Phi_0e^{-i\theta}|^2 + \frac{1}{2}|(\d_\mu - ie{\cal A}_\mu)\chi_0 e^{-i\phi}|^2 \right.\cr
&&\left.+ \epsilon^{\mu\nu\rho}\left(\frac{1}{e}{\cal A}_\mu \d_\nu \tilde b_\rho + a_\mu \d_\nu\tilde b_\rho\right) + V(\phi_0^2) + V(\chi_0^2)\right]\;,
\label{dualityaction}
\eea
which reduces to our action (\ref{action}) by putting $\chi_0=0={\cal A}_\mu$ and renaming $\tilde b_\mu\rightarrow A_\mu$, 
can also be embedded in the ABJM model, by 
\bea
&&A_\mu=a_\mu^{(1)} \mathbf{1}_{N\times N}\,,\cr
&&\hat A_\mu=\hat a_\mu ^{(1)}\mathbf{1}_{N\times N}\,,\cr
&&Q^1=\phi G^1_{N\times N}\,,\cr
&&Q^2=\phi G^2_{N\times N}\,,\cr
&& R^\a=0\,,\label{embedABJM}
\eea
and the identifications 
\be
\Phi\rightarrow \phi, \;\;\;a_\mu\rightarrow a^{(1)}_\mu-\hat a^{(1)}_\mu,\;\;\; (A_\mu=){\tilde b}_\mu\rightarrow
a^{(1)}_\mu+\hat a^{(1)}_\mu.
\ee
Note that in \cite{Murugan:2014sfa}, since one wanted to obtain the whole action (\ref{dualityaction}), the embedding (\ref{embedABJM}) was 
only for a $N/2\times N/2$ subspace, and the other half of the fields, $\chi_0$ and ${\cal A}$, now put to zero, was obtained from another 
$N/2\times N/2$ subspace, together with an identification. But in our case, we don't need to do that. 

{\bf Obtaining source by duality}

Another possible answer is that we can use the Mukhi-Papageorgakis Higgs mechanism \cite{Mukhi:2008ux}
 in the form in \cite{Murugan:2014sfa} on the
action (\ref{dualityaction}) (embeddable in the ABJM model), as follows. 

First, expand 
\be
\chi=\chi_0 e^{-i\phi}=(\langle \chi_0\rangle+\delta \chi_0)e^{-i\theta \phi};\;\;\; \delta \phi=\delta \phi_{\rm smooth}+\delta\phi_{\rm vortex}.
\ee
Then the Chern-Simons gauge field ${\cal A}_\mu$ eats the scalar $\delta\phi$ and becomes Maxwell (just like in the regular Higgs mechanism
the Maxwell field eats a scalar and becomes massive) through the shift (first equality) and duality (second equality) relation
\be
e{\cal A}_\mu+\d_\mu \phi_{\rm smooth} +\d_\mu \phi_{\rm vortex} \equiv e {\cal A}'_\mu =\frac{1}{e^2\langle\chi_0\rangle^2}
{\epsilon_\mu}^{\nu\rho}\d_\nu A_\rho\,.\label{calA}
\ee
Defining as usual $F_{\mu\nu}=\d_\mu A_\nu-\d_\nu A_\mu$, the action becomes
\bea
S&=&-\int d^3x\ \left[\frac{1}{2}|(\d_\mu - iea_\mu)\Phi_0e^{-i\theta}|^2 + \frac{1}{2}(\d_\mu \delta \chi_0)^2
+ \frac{1}{4e^2\langle\chi_0\rangle^2}F_{\mu\nu}F^{\mu\nu} + \epsilon^{\mu\nu\rho}a_\mu \d_\nu A_\rho\right.\cr
&&\left.-\frac{2\pi}{e}A_\mu  j^\mu_{\rm vortex}(t)+V(\Phi_0^2)+V(\chi_0^2)\right]\;,
\eea
where $j^\mu_{\rm vortex}$ is associated with $\phi_{\rm vortex}$. Then, first putting $\delta\chi=0$, and considering that the Maxwell term for 
$A_\mu$ can be neglected at low energies with respect to the Chern-Simons term, we arrive at our starting action (\ref{action}), with $k=1$ and 
a rescaling $A_\mu\rightarrow A_\mu/(2\pi)$. 

{\bf Action from gravity dual of FQHE}

The ABJM model is dual to string theory in $AdS_4\times \mathbb{CP}^3$ background, but we can also consider a gravity dual construction 
that gives directly the FQHE, like in one in \cite{Hikida:2009tp}.
In this case, the starting action (\ref{action}) (except for the source term) is obtained by definition, but we want to stress here that there 
is a well defined gravity dual model. 

One considers probe D4-branes wrapping 2-cycles (fractional D2-branes) in the gravity dual. Then the WZ term
\be
2\pi^2 T_4\int_{\mathbb{R}^{1,2} \times \mathbb{CP}^1} C_1\wedge F\wedge F=
2\pi^2 T_4\left[\int_{\mathbb{CP}^1} dC_1\right]\int_{\mathbb{R}^{1,2}} A\wedge dA
\ee
becomes one of the terms in the FQHE effective action, $\sim a\wedge da$ ($a$ the statistical gauge field). 
The second is obtained by an having an additional background 3-form field $C_3$ of the type
\be
C_3=\frac{4\pi k}{R^3}A_{ext}\wedge \omega\;.
\ee
In this case, the WZ term
\be
\frac{1}{(2\pi)^4}\int_{\mathbb{R}^{1,2}\times \mathbb{CP}^3}2\pi F\wedge C_3=\left[\frac{1}{(2\pi)^3}\int_{\mathbb{CP}^1}
\frac{8\pi^2k}{R^3}\omega\right]\frac{1}{2\pi}\int_{\mathbb{R}^{1,2}}A_{ext}\wedge F
\ee
gives rise to the second term in the FQHE effective action, $A\wedge da$ ($A_{ext}\rightarrow A$ is the electromagnetic field and $A\rightarrow a$ is the 
statistical gauge field).

{\bf Action from SYM-CS theories}

Another interesting model in 2+1 dimensions is ${\cal N}=8$ (maximal) $SU(N)$ SYM with Chern-Simons terms at level $k$. Such a model, 
for some particular CS terms, was found to be dual to a massive type IIA solution by Guarino, Jafferis and Varela (GJV) in \cite{Guarino:2015jca}.

In this case the $SU(N)$ Maxwell term would be subleading at low energies with respect to the Chern-Simons terms. The SYM will also have 
complex scalars coupled to the $SU(N)$ gauge field, and in general a sextic potential for these scalars. From the action (\ref{action}) 
the first and the last term are obtained, and from $S'$ we can obtain easily the CS term. The source term(s) have to be added by hand. 
But moreover, now we would need to also introduce by hand the Chern-Simons term coupling to an external gauge field.

\section{Discussion and conclusions}

In this paper we have studied vortex soliton solutions of three dimensional scalar gauge theories with sources. The presence of Chern-Simons 
terms and sources can make the vortex an anyon, and the presence of the Chern-Simons term mixing the statistical and electromagnetic gauge fields 
can make it relevant to the Fractional Quantum Hall Effect. We have seen that there are vortex solutions that have the statistical gauge field 
$a_\mu=\d_\mu\theta/e$, just like in the case of the anyon construction in section 2.1, and for which the equations of motion reduce to the 
radial scalar equation of motion, the same as the classical motion in an inverted potential $V_m(X)=-V(X)$, with $r\rightarrow t$ and $|\Phi(r)|\rightarrow X(t)$. 

We have classified all possible renormalizable potentials in terms of whether they give vortex solutions, using the classical mechanics analogy. We have 
shown what are the possible usual solitons, with finite total energy, $|\Phi(r\rightarrow 0)|\rightarrow 0$ and $|\Phi(r)|$ extending all the way to infinity, 
where it has zero potential. We have also analyzed "compacton" vortices, that end at a finite $r_0$ away from the vortex core.
A very interesting case of nonrenormalizable potential was considered with the intent of obtaining Laughlin's wavefunction as a vortex solution. 
We have shown that this is only possible if we consider the approximation where we have different UV ($r\rightarrow 0$) and IR ($r\rightarrow \infty$) 
potentials $V(|\Phi|)$. We obtained the Laughlin wavefunction in the case $m=1/\nu\rightarrow \infty$ natuarally, where the small $|\Phi|$ potential 
in the UV is dominated by the mass term $m^2|\Phi|^2/2$, and the small  $|\Phi|$ potential in the IR is dominated by a log-corrected mass term, 
$-m^2|\Phi|^2\ln |\Phi|^2$, possibly the effect of a (one-loop) Wilsonian effective action calculation. 

Note that one possible interpretation of the presence of a source term in the equations of motion for the vortices would be if the 2+1 dimensional system
described by the above construction sits at a boundary of a 3+1 dimensional material which has vortex strings. Then the 
dynamics at this boundary can be a purely 2+1 dimensional one (independent of 3+1 dimensions), with the exception of the sources coming from the 
endpoints of the vortex strings (whose flux is uncompensated, so acts as a 2+1 dimensional source). Such a situation could arise in 
specific constructions of a fractional quantum Hall system. 

Vortices in nonabelian theories were also considered, though we only considered the generic case of embedding the abelian vortices in them, not 
looking for specific nonabelian vortices, which would depend on the specific nonabelian model. We leave this interesting extension for the future. 

A nonrelativistic limit of our models gave an interesting vortex solution, which we wrote explicitly. It is a solution of the compacton type, ending at 
a finite $r_0$, and also similar to a "peakon" type, by having a discontinuity in the first derivative at $r_0$, yet still being a solution of the equations of motion. 
The solution is characterized by a vortex number, but also an {\em independent} magnetic flux number. 

In a search for concrete applications of our generic set-up to specific fundamental models, like the ones coming from string theory (and relevant for 
the AdS/CFT correspondence), we have found that we can embed our action in the ABJM model, in the SYM-CS model dual to the GJV solution, as
well as have a specific string theory construction that is relevant for the Fractional Quantum Hall Effect. 

We believe that the analysis in this paper has only begun the task of exploring the physical consequences of our vortex solutions in the FQHE. 
The role of the vortex solutions in describing the dynamics, for example in deriving from first principles the wavefunctions of the FQHE states
(Laughlin and excited ones), as well as the dynamics of vortices, described by the effective Lagrangean in the moduli space approximation, 
remain to be understood in further work.

\section*{Acknowledgements}

We would like to thank Chrysostomos Kalousios for collaboration at the initial stages of this project. We would also like to thank Jeff Murugan and 
Carlos N\'{u}\~nez for useful comments and a careful reading of the manuscript.
The work of HN is supported in part by CNPq grant 301219/2010-9 and FAPESP grant 2014/18634-9. 
The work of FR is supported by grant 2012/05451-8. F.R. would also like to thank the DFI-FCFM of Universidad de Chile for hospitality during part of this work.

\bibliography{solitonFQHE}
\bibliographystyle{utphys}

\end{document}